\documentclass[a4paper,11pt]{article}
\pdfoutput=1 

\usepackage{jheppub} 

\usepackage{amsmath,amsfonts,amssymb,bbm,mathrsfs,mathtools,amsthm}
\usepackage{graphicx}
\usepackage{subcaption}
\usepackage{slashed,cancel}
\usepackage{bm}
\usepackage{booktabs}
\usepackage{braket}
\usepackage{bbold}
\usepackage{multirow}
\usepackage{tikz}
\usepackage{tikz-feynman}
\usepackage[force]{feynmp-auto}
\usepackage{orcidlink} 
\usepackage[normalem]{ulem} 
\usepackage[capitalise]{cleveref}
\usepackage[T1]{fontenc} 
\usepackage{overpic}

\newcommand{\be}{\begin{equation}}
\newcommand{\ee}{\end{equation}}
\newcommand{\bea}{\begin{eqnarray}}
\newcommand{\eea}{\end{eqnarray}}
\newcommand{\ba}{\begin{aligned}}
\newcommand{\ea}{\end{aligned}}

\newcommand{\figref}[1]{Fig.~\ref{#1}}

\newcommand{\secref}[1]{Section~\ref{#1}}

\newcommand{\tabref}[1]{Table~\ref{#1}}

\title{\boldmath nuSTORM as a Precision Probe of the Standard Model and New Physics}

\preprint{IPPP/25/54, IFT-UAM/CSIC-25-94}

\author[a]{Jack Franklin~\orcidlink{}\,,}
\author[b]{Rohan Kamath~\orcidlink{0009-0005-5683-0614}\,,}
\author[c]{Dhruv Pasari~\orcidlink{0009-0007-1283-1492}\,,}
\author[d]{Yuber F. Perez-Gonzalez~\orcidlink{0000-0002-2020-7223
}\,,}
\author[c]{Jessica Turner~\orcidlink{0000-0002-9679-5252}\,,}
\author[b]{Maria Athina Vogiatzi~\orcidlink{0000-0000-0000-0000}}

\affiliation[a]{The Institute of Computing for Climate Science, Centre for Mathematical Sciences, Wilberforce Road, Cambridge, CB3 0WA, U.K.}
\affiliation[b]{Physics Department, Blackett Laboratory, Imperial College London, London, SW7 2AZ, U.K.}
\affiliation[c]{Institute for Particle Physics Phenomenology, Department of Physics, Durham University, Durham, DH1 3LE, U.K.}
\affiliation[d]{Departamento de Física Teórica and Instituto de Física Teórica UAM/CSIC, Universidad Autónoma de Madrid, Cantoblanco, 28049 Madrid, Spain}

\emailAdd{jdf58@cam.ac.uk}
\emailAdd{rohan.kamath16@imperial.ac.uk}
\emailAdd{dhruv.pasari@durham.ac.uk}
\emailAdd{yuber.perez@uam.es}
\emailAdd{jessica.turner@durham.ac.uk}
\emailAdd{m.vogiatzi24@imperial.ac.uk}

\abstract{The Neutrinos from Stored Muons (nuSTORM) facility will generate neutrino beams from both muon and meson decays in a storage ring, providing a neutrino flux known to the percent level. This unprecedented precision enables a rich physics programme, including high-precision tests of the Standard Model and searches for new phenomena. In this paper we demonstrate nuSTORM's sensitivity to key Standard Model processes such as, measurements of the weak mixing angle at low $Q^2$ and the rare process of neutrino trident production. We also show its powerful reach for a diverse range of beyond-the-Standard-Model scenarios, including eV-scale sterile neutrinos, Kaluza-Klein excitations from large extra dimensions and lepton flavour violation. Furthermore, nuSTORM can place significant constraints on heavy QCD axions and other axion-like particles produced in rare kaon decays. These capabilities establish nuSTORM as a powerful and complementary probe to long baseline experiments and collider searches.}


\begin{document}
\maketitle
\flushbottom

\section{Introduction}
\label{sec:intro}
The Neutrinos from Stored Muons (nuSTORM) facility is a proposed next-generation experiment designed to deliver the most precisely characterised neutrino beams ever constructed~\cite{nuSTORM:2025tph}. By generating neutrinos from the decay of muons circulating in a storage ring, nuSTORM will facilitate high-precision studies of neutrino interactions, provide critical support for the broader neutrino physics programme and open a new window to physics Beyond the Standard Model (BSM).

The motivation for nuSTORM is threefold, addressing key challenges at both the experimental and theoretical frontiers. First, upcoming long baseline oscillation experiments, such as DUNE~\cite{DUNE:2020ypp} and Hyper-Kamiokande~\cite{Hyper-KamiokandeProto-:2015xww}, require percent-level precision on neutrino cross-sections to achieve their full potential, particularly in the search for leptonic CP violation. Currently, their expected largest systematic uncertainty is our limited understanding of neutrino-nucleus interactions that are crucial for appearance measurements. nuSTORM confronts this challenge directly by producing beams of $\nu_e$ and $\nu_\mu$, with precisely known flavour compositions, and energy spectra characterised by flux normalisation understood at the sub-percent level.  Second, the facility provides a unique platform for advancing theoretical frontiers: by delivering high-precision neutrino data, along with access to variable muon momentum, nuSTORM enables stringent tests of the Standard Model and opens broad sensitivity to new phenomena. While providing cross-section data is a primary objective of the nuSTORM programme, this work focuses on the facility's potential for other high-impact Standard Model measurements and its reach in exploring physics beyond the Standard Model. Finally, nuSTORM serves as a stepping stone toward a muon collider, as a testbed for novel technologies involved in the production, monitoring and storage of high-flux and muon beams.

This physics programme includes unique probes of the Standard Model, such as precision measurements of the weak mixing angle and the rare process of neutrino trident production, which both test the electroweak gauge structure. In parallel, nuSTORM's exceptional control over beam properties makes it a powerful and versatile instrument for BSM searches. The beams with high-intensity, and known flavour composition are ideal for a diverse range of explorations. For instance, short-baseline oscillation patterns can be scrutinised for evidence of eV-scale sterile neutrinos, while subtle spectral distortions may reveal Kaluza-Klein excitations predicted by models with large extra dimensions. The clean experimental environment also enables sensitive searches for lepton flavour violation in meson decays and can place stringent constraints on heavy QCD axions and other axion-like particles produced in rare kaon decays.

This paper provides a wide ranging exploration of nuSTORM’s physics. 
Section~\ref{sec:setup} introduces the experimental design, describing the neutrino beam simulation, detector specifications and the assumptions underlying our phenomenological analyses. 
We then turn to physics studies: Standard Model measurements are presented in Section~\ref{sec:SM}, followed by an investigation of nuSTORM’s sensitivity to new physics, including sterile neutrinos (Section~\ref{sec:steriles}), large extra dimensions (Section~\ref{sec:LED}), lepton flavour violation (Section~\ref{sec:LFV}), neutrino trident production (Section~\ref{sec:trident}) and axion-like particles (Section~\ref{sec:axion}). 
Concluding remarks are provided in Section~\ref{sec:conclusion}.

\section{The nuSTORM Experimental Setup}\label{sec:setup}
In this section, we detail the key features of the nuSTORM facility, including the accelerator design, beam properties and the detector configuration assumed for our analysis.

\subsection{Design Overview}
\begin{figure}[t!]
    \centering
    \begin{overpic}[width=\columnwidth]{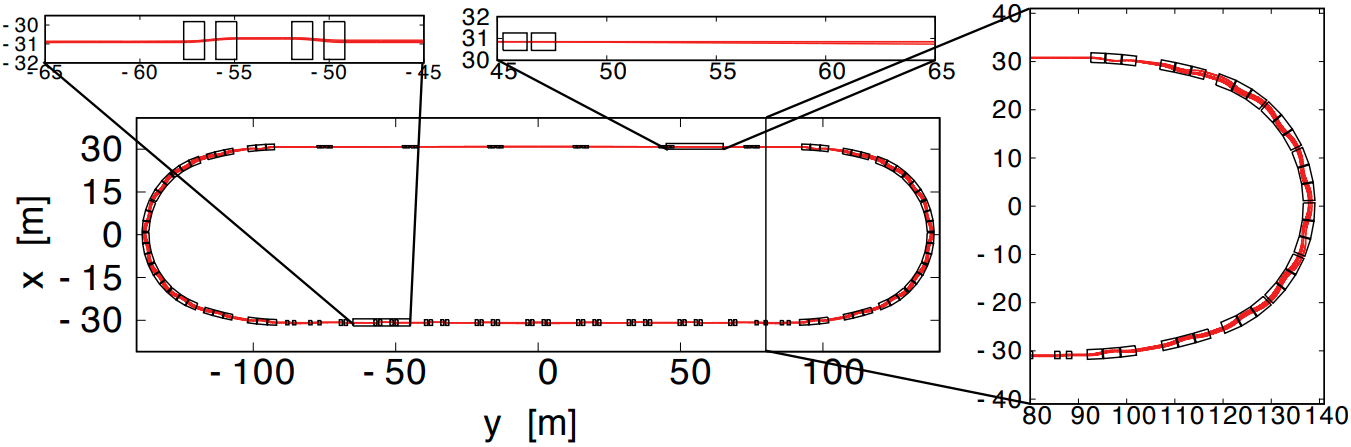} 
        \put(28, 19.){\small Production straight} 
        \put(17, 19.){\small OCS} 

        \put(31, 10.5){\small  Return straight}
    \end{overpic}
    \vspace{-0.2cm}
    \caption{Schematic of the nuSTORM Experiment: Pions generated from a target and focusing horn are injected into a racetrack-shaped muon storage ring with a detector placed downstream of the production straight. \cite{nuSTORM:2022div}}
    \label{fig:nustorm_schematic}

\end{figure}

nuSTORM is a 616-meter circumference racetrack-shaped muon storage ring, designed to store muons between $1$ - $6$ GeV/c with a momentum acceptance of $\pm 16\%$. The decay of these stored muons yields equal production rates of muon and electron neutrinos with well characterised flux properties.  A focusing horn system collects charge selected pions from proton target interactions. The pions of momentum $p_\pi$ are transported through the transfer line and injected into the production straight section of the storage ring shown in Fig.\ref{fig:nustorm_schematic} at the Orbit Combination Section (OCS), with a pion momentum acceptance of $\pm 10\%$. The current lattice design captures muons with central momentum $p_\mu = 0.76 p_\pi$ for circulation, ensuring momentum separation to prevent pions from circulating in the ring. A 5m $\times$ 5m detector face 50m downstream observes an initial neutrino flash of muon neutrinos from those captured-pion decays, followed by periodic signals of muon and electron neutrinos from muon decays as the stored muons circulate through the ring. The neutrino flavours can be controlled by injecting $\pi^+$ or $\pi^-$ through forward and reverse horn current configurations. Here, pions decay to produce muons that are stored in the ring, while undecayed pions are dumped downstream of the production straight. By varying the pion injection (and hence muon storage energies), this design enables comprehensive coverage of the neutrino energy spectrum relevant to current and future long baseline experiments. It can also probe the full range of neutrino interaction modes, from the charged-current quasi-elastic (CCQE) to the deep inelastic scattering (DIS) dominated regimes.

nuSTORM requires high dynamical acceptance, with muons in stable orbit in a large region of phase space. It also requires a large momentum acceptance and efficient muon capture. A hybrid lattice design addresses these requirements. The production straight employs a conventional room-temperature focusing-defocusing quadrupole magnetic lattice. This structure minimises dispersion and ensure that the injected pions and recirculating muons share the same orbital path, allowing for maximum muon capture efficiency. The remainder of the lattice utilises Fixed Field Alternating gradient (FFA) magnets to achieve $\pm 16\%$ momentum acceptance with minimum chromaticity. Superconducting high-field combined magnets in the arcs minimise arc length to maximise muon time in the straight sections, optimising neutrino production toward the detectors. The return straight incorporates room-temperature combined-function magnets.

\subsection{Simulations}
The nuSTORM design and simulation studies are based on a reference configuration with the facility located at CERN, leveraging the existing proton driver infrastructure. Simulations of the target and horn system model a 100 GeV proton beam from the Super Proton Synchrotron (SPS) impinging on an inconel target. These simulations have been performed using FLUKA \cite{Battistoni2015,Ahdida2022}, a general purpose Monte Carlo tool for particle transport and interaction calculations with applications spanning accelerator physics and radiotherapy. These simulations provide output particle distributions on the downstream end of the horn for subsequent simulation studies. 

Comprehensive simulations of muon production and the resultant neutrino fluxes from FLUKA-generated horn output distributions at nuSTORM have been carried out using two complementary approaches. Beam Delivery Simulation (BDSIM) \cite{Nevay2020} provides detailed particle tracking and decays of input pion and kaon distributions and their decay products, incorporating models of the transfer line and production straight magnets, with ongoing development to include FFA magnet capabilities for full-ring simulations. In parallel, the Python-based bespoke nuSTORM SIMulator (nuSIM) framework offers a faster alternative by stochastically sampling pion decay positions, applying relativistic decay kinematics along with lattice acceptance cuts, and similarly propagating the muons and their subsequent decays. Neutrino fluxes are then estimated by counting the resulting neutrinos from the respective decays passing through a reference detector plane, enabling efficient studies of beam production characteristics while detailed tracking models continue to be developed. Detailed discussions of the above simulations have been presented in Alvarez Ruso et. al \cite{nuSTORM:2022div}.

\begin{figure}[t!]
  \centering
  \includegraphics[width=\linewidth]{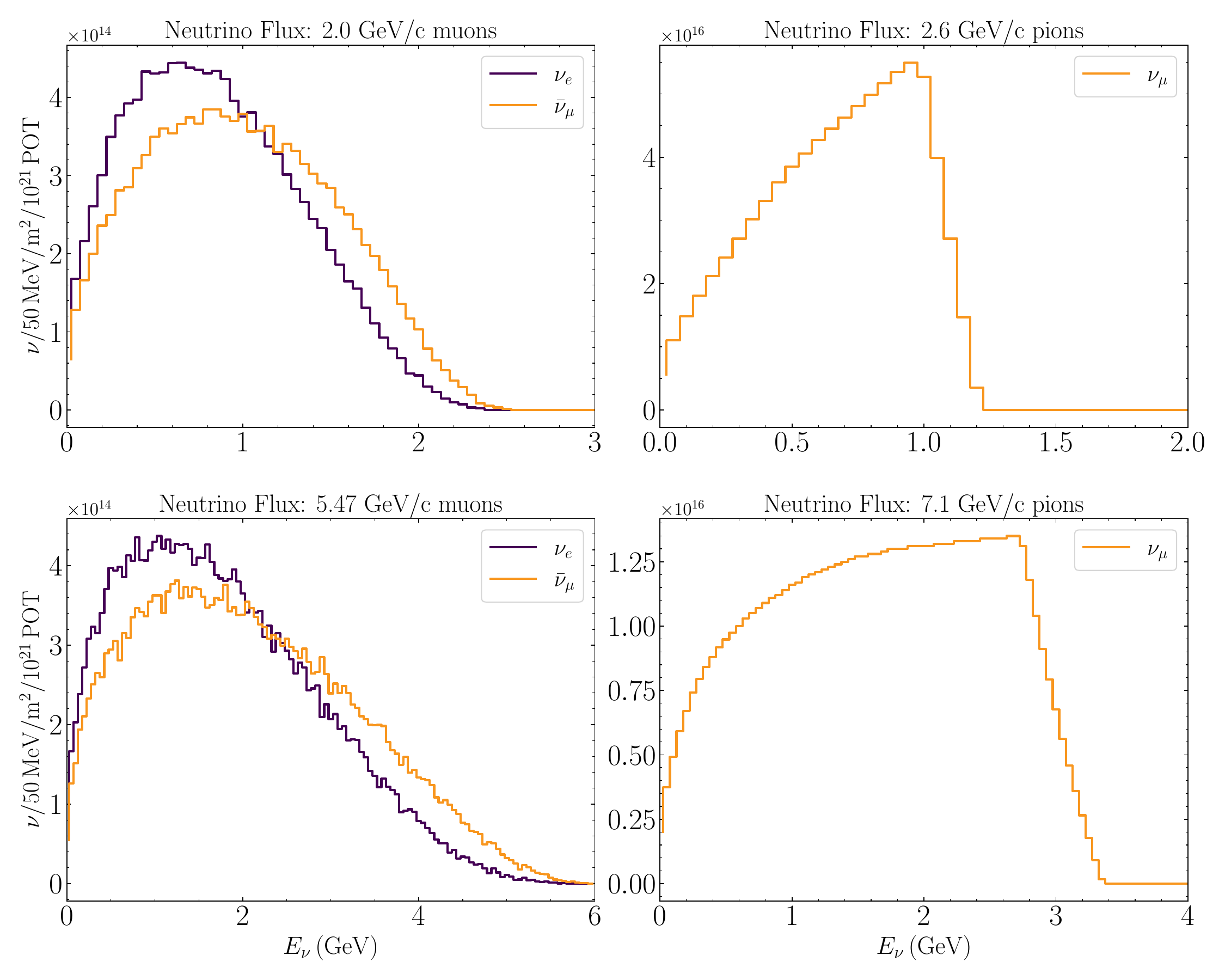}
  \caption{Neutrino fluxes for four representative nuSTORM running conditions given four different muon beam energies.}
  \label{fig:nustorm_flux}
\end{figure}

\subsection{Detector Specifications and Analysis Assumptions}

A key advantage of nuSTORM is its ability to deliver a neutrino beam with an exceptionally well-understood flux, which is the dominant source of systematic uncertainty in most neutrino experiments. This precision is possible because the neutrinos originate from muon decay ($\mu \to e \nu_e \nu_\mu$), a purely leptonic process that is precisely calculable within the Standard Model. By accurately monitoring the muon intensity in the storage ring, the resulting neutrino flux, including its flavour composition and energy spectrum, can be predicted to the sub-percent level.

This paper presents a phenomenological study of nuSTORM's physics reach. To isolate the impact of statistical uncertainties and the intrinsic potential of the facility, we assume an idealised detector with 100$\%$ detection efficiency and perfect particle identification. We note that the assumption of $100\%$ detection efficiency is optimistic and based on the assumption of highly efficient LArTPC detectors. Naturally, including more realistic detector efficiency would reduce the overall sensitivity to each physics search.  For our analysis, we consider a detector with a fiducial mass of approximately 100 tonne, similar in scale to the ProtoDUNE Liquid Argon Time Projection Chambers (LArTPC) at CERN. 
Throughout this work, we assume a total integrated exposure of {$10^{21}$ Protons on Target (POT)}, which corresponds to approximately 10 years of operation at nuSTORM's proposed beam power. It must be noted that each phenomenological analysis is a standalone analysis, and hence we assume an optimistic framework where all $10^{21}$ POT are employed towards the relevant energy mode.

To calculate the expected event rates, neutrino interaction cross-sections are taken from the GENIE event generator with tune \href{https://genie-mc.github.io/tunes.html}{\texttt{G1810a0211b-k250-e1000}}
 \cite{Andreopoulos:2015wxa}. To maximise the available statistics, our analysis combines events from both the initial, high-energy ``pion flash''  neutrinos (predominantly $\nu_\mu$) and the subsequent ``signal'' neutrinos produced from the decays of the stored muons. The fluxes of the pion flash and muon signal neutrinos are shown in \figref{fig:nustorm_flux} for relevant beam energies.

\subsection{Number of Events Estimation}\label{sec:event_estimation}
The expected number of events at nuSTORM is estimated using the standard relation  
\begin{equation}
\label{eq:Nevents_estimate}
N_\mathrm{events} = N_\mathrm{POT} \times N_\mathrm{targets} \int \frac{d\Phi(E_\nu)}{dE_\nu} \, \sigma(E_\nu) \, dE_\nu \,,
\end{equation}
where $N_\mathrm{POT}$ is the number of protons on target, $N_\mathrm{targets}$ the number of scattering centres in the detector, $d\Phi/dE_\nu$ the differential neutrino flux (see \figref{fig:nustorm_flux}) and $\sigma(E_\nu)$ the energy-dependent interaction cross-section. The integration is performed over the full neutrino energy range relevant to the analysis.  

For oscillation studies, fluxes from low-energy muon decays are typically employed. This choice reflects the $L/E_\nu$ dependence of the oscillation probability: at a fixed baseline, lower neutrino energies enhance sensitivity to oscillation parameters. In this case, Eq.~\ref{eq:Nevents_estimate} is modified by an additional factor corresponding to the oscillation probability. 
In contrast, for rare processes or precision measurements of Standard Model parameters, fluxes from higher-energy muon decays are generally preferred. Here the increasing cross-section with energy yields higher event rates and improved statistical precision, as in studies of neutrino trident production, weak mixing angle measurements and rare kaon decay channels. Unless otherwise specified, a 100\% detector efficiency is assumed throughout. That is, all final states produced within the detector volume are taken to be detected without loss. This idealised assumption provides a common baseline for sensitivity studies and enables straightforward comparison across physics channels and experimental configurations. The resulting event yields serve as input to the statistical analysis described below.

\subsection{Statistical Framework and Test Statistic}\label{sec:statistical_framework}

To estimate the sensitivity of nuSTORM to Standard Model precision observables ($e.g.$, the weak mixing angle) and to Beyond Standard Model parameters ($e.g.$, sterile neutrinos or large extra dimensions), we adopt a general statistical framework based on a binned log-likelihood $\chi^2$ comparison between the theoretical prediction and the Standard Model expectation. The test statistic is defined as
\begin{equation}
\label{eq:chi_squared_general}
\Delta \chi^2 = \min_{\eta} \left[
\underbrace{
2 \sum_{\mathrm{channels}} \sum_{i}^{\mathrm{bins}} 
\left\{
N_i^{\mathrm{theo}}(\vec{\theta},\eta) - N_i^{\mathrm{std}} 
+ N_i^{\mathrm{std}} \ln \frac{N_i^{\mathrm{std}}}{N_i^{\mathrm{theo}}(\vec{\theta},\eta)}
\right\}
}_{\text{statistical term}}
+ 
\underbrace{
\sum_k \frac{\eta_k^2}{\sigma_k^2}
}_{\text{systematic penalty}}
\right]\,,
\end{equation}
where $N_i^{\mathrm{theo}}(\vec{\theta},\eta)$ is the predicted number of events in the $i$-th bin for a given theory with model parameter(s) $\vec{\theta}$ ($e.g.$, neutrino oscillation parameters, weak mixing angle, or new-physics coupling) and $N_i^{\mathrm{std}}$ is the Standard Model prediction. The sum over channels includes all neutrino flavours and interaction modes relevant to the analysis at nuSTORM are $\nu_e \rightarrow \nu_e$ and $\bar{\nu}_{\mu}\rightarrow\bar{\nu}_{\mu}$ from muon decays and $\nu_{\mu} \rightarrow \nu_{\mu}$ from the prompt pion-decay flux.

Systematic uncertainties are incorporated as nuisance parameters $\eta_k$, each with a prior $\sigma_k = 1\%$~\cite{Adey:2015iha, nuSTORM:2022div}. These parameters are allowed to vary conservatively within $\pm 5\%$ during the minimisation. Since the dominant systematic uncertainties from flux prediction and cross-section modelling are individually strongly correlated across all channels, a common nuisance parameter is adopted.
 This framework allows for a consistent treatment across different physics cases, providing a robust estimate of the experimental sensitivity while accounting for correlated systematic effects.
 The key assumptions of the experiments used for comparision against nuSTORM are presented in \tabref{tab:exp-summary}.

\begin{table}[t!]
    \centering
    \renewcommand{\arraystretch}{1.15}
    \resizebox{\textwidth}{!}{%
    \begin{tabular}{lccccc}
        \toprule
        \textbf{Experiment} & \textbf{Baseline} & \textbf{Exposure / POT} & \textbf{Mass/Volume} & \textbf{Energy} \\
        \midrule
        nuSTORM                           & 50 m        & $10^{21}$              & 100 t  & $0-6$ GeV \\
        DUNE (Near)~\cite{DUNE:2020ypp}   & 574 m       & $1.47\times10^{22}$/yr & 50 t  & $0-40$ GeV \\
        DUNE (Far)~\cite{DUNE:2020ypp}    & 1300 km     & $1.47\times10^{22}$/yr & 40 kt  & $0-20$ GeV \\
        SBND~\cite{Bonesini:2022pwy}      & 110 m       & $6.6\times10^{20}$  & 112 t  & $0-3$ GeV \\
        MicroBooNE~\cite{Bonesini:2022pwy}& 470 m       & $1.32\times10^{21}$  & 85 t   & $0-3$ GeV \\
        ICARUS~\cite{Bonesini:2022pwy}    & 600 m       & $6.6\times10^{20}$    & 476 t  & $0-3$ GeV \\
        MINOS/MINOS+~\cite{Forero:2022skg}& 735 km      & $16.36\times10^{20}$ (Combined)  & 5.4 kt     & $0-40$ GeV \\
        Daya Bay~\cite{DayaBay:2018yms}   & 0.36–1.9 km & $17\,\mathrm{GW_{th}}$  & 20 t$\times$8 detectors  & $\sim 3-4$ MeV \\
        FASER2~\cite{FASER:2019aik}       & 480 m       & $1.1\times10^{16}$& $1\mathrm{m}\times1\mathrm{m}\times7\mathrm{m}$ & $\sim\mathcal{O}( \mathrm{GeV\, -TeV})$ \\
        \bottomrule
    \end{tabular}%
    }
    \caption{Key parameters of representative neutrino experiments relevant for comparison with nuSTORM.}
    \label{tab:exp-summary}
\end{table}

\section{Standard Model Measurements at nuSTORM}\label{sec:SM}
In the following section, we discuss two precision tests of the Standard Model within the sensitivity reach of nuSTORM, namely the weak mixing angle and neutrino trident scattering. Measurements such as these serve as indirect probes of new physics. 

\subsection{Weak Mixing Angle}\label{sec:weak mixing}
The weak mixing angle (or Weinberg angle), denoted $\theta_W$, is a fundamental parameter of the Standard Model's electroweak sector. It relates the SU(2)$_L$ and U(1)$_Y$ gauge couplings, $g$ and $g'$, via the renormalisation scale-dependent quantity
\begin{equation}
\sin^2\theta_W(\mu) \equiv \frac{g'^2(\mu)}{g^2(\mu) + g'^2(\mu)},
\end{equation}
where $\mu$ denotes the renormalisation scale.
Precise measurements of $\sin^2\theta_W$ have been performed across a wide range of energy scales and processes, including atomic parity violation~\cite{Dzuba:2012kx, Roberts:2014bka}, M{\o}ller scattering, coherent elastic neutrino nucleus scattering(CE$\nu$NS)~\cite{DeRomeri:2022twg} and neutrino-electron scattering~\cite{deGouvea:2019wav}. A detailed study on these methods is presented in~\cite{ParticleDataGroup:2022pth}. One notable result comes from the NuTeV experiment, which reported
\begin{equation}
\sin^2\theta_W = 0.2407 \pm 0.0016 \quad (\overline{\text{MS}}, \,\langle Q \rangle \approx 4.5~\mathrm{GeV})\,,
\end{equation}
a value that deviated from the global electroweak fit by approximately $3\sigma$~\cite{NuTeV:2001whx}. While this anomaly could hint at new physics, it might also stem from uncharacterised nuclear effects or uncertainties in parton distribution functions~\cite{Bentz:2009yy,Brodsky:2004qa,Dobrescu:2003ta}. Resolving this tension requires new, precise measurements in theoretically clean environments.
\begin{figure}[t!]
    \centering
    \includegraphics[width=0.85\linewidth]{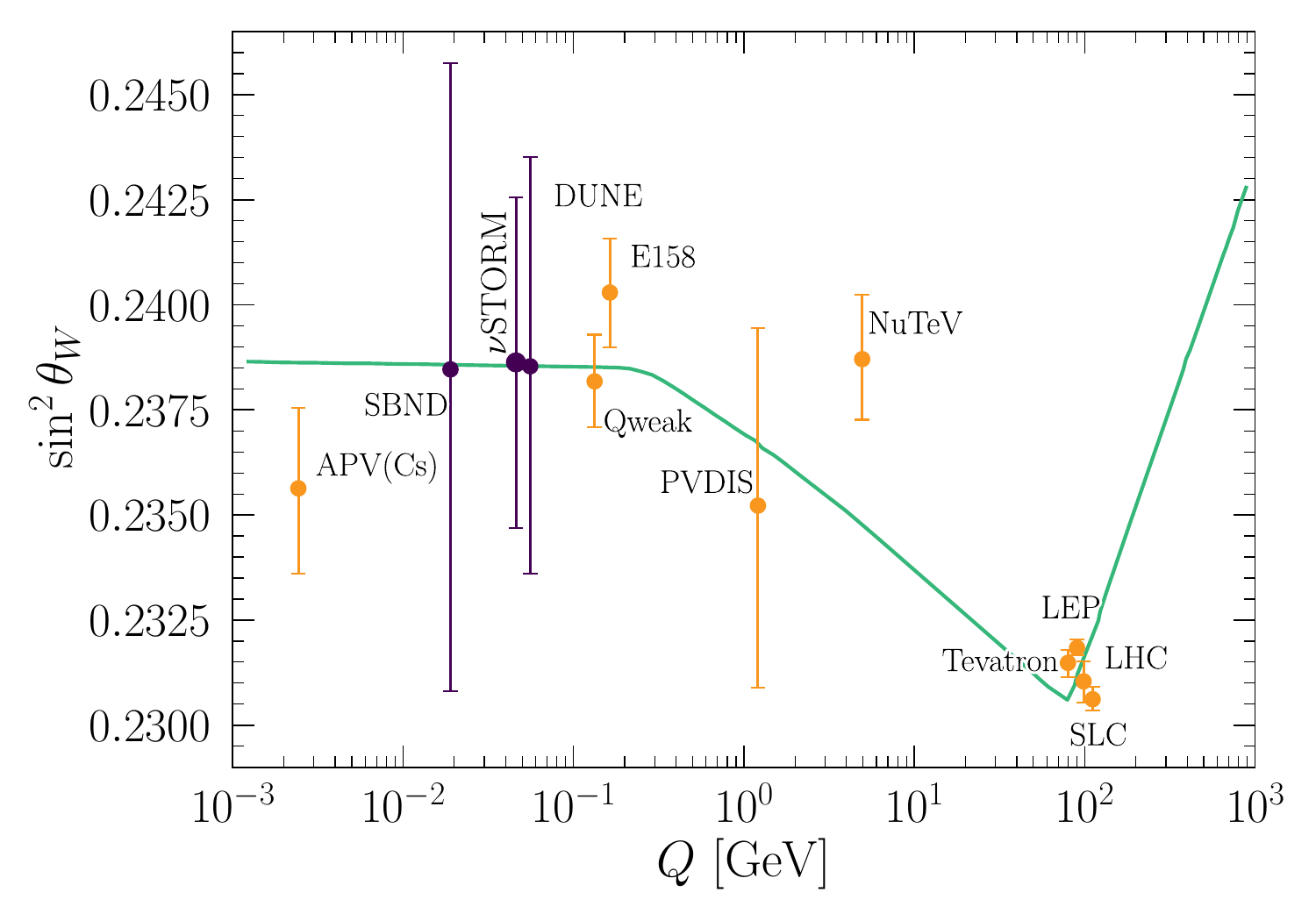}
    \caption{Running of the weak mixing angle $\sin^2\theta_W$ with the momentum transfer $Q$ (green curve), compared against existing measurements (orange points). The projected sensitivities from nuSTORM, SBND~\cite{Alves:2024twb}, and DUNE-PRISM~\cite{deGouvea:2019wav} are shown in purple. The DUNE-PRISM point corresponds to an electron angular resolution of $\sigma_\theta = 1^\circ$, and SBND-PRISM corresponds to an optimistic benchmark of $10^{22}$ POT with $5\%$ correlated systematics. The data points from Tevatron, SLC, and LHC have been slightly shifted from $Q= M_Z$ for improved visibility.}
    \label{fig:chi_square_contour_weak}
\end{figure}
The theoretically cleanest channel for measuring $\sin^2\theta_W$ is neutrino-electron scattering, as it is free from hadronic and nuclear uncertainties. Its primary limitation is the small cross-section, which demands an intense neutrino source. For $\nu_\mu e^- \to \nu_\mu e^-$ scattering, the differential cross-section depends on the electron recoil energy $E_R$ and incident neutrino energy $E_\nu$. It is proportional to the SM couplings of the $Z$ boson to the electron:
\begin{align}
g_V &= -\frac{1}{2} + 2 \sin^2\theta_W\,, \\
g_A &= -\frac{1}{2}\,.
\end{align}
The general differential cross-section is given by:
\begin{equation}
\frac{d\sigma}{dE_R} = \frac{2 G_F^2 m_e}{\pi} 
\left[
g_1^2 + g_2^2 \left( 1 - \frac{E_R}{E_\nu} \right)^2 - g_1 g_2 \frac{m_e E_R}{E_\nu^2}
\right]\,,
\end{equation}
where the terms $g_1$ and $g_2$ are functions of $g_V$ and $g_A$~\cite{deGouvea:2019wav}. For the specific channels of interest, the cross-sections can be expressed as:
\begin{equation}
\frac{d\sigma_{\nu_{\mu}}}{dE_R} \propto \left(\frac{1}{4}-\sin^2\theta_W\right) + \sin^4\theta_W\left(2-\frac{2E_R}{E_{\nu}}+\frac{E_R^2}{E_{\nu}^2}\right)\,,
\end{equation}

\begin{equation}
\frac{d\sigma_{\nu_{e}}}{dE_R} \propto \left(\frac{1}{4}+\sin^2\theta_W\right) + \sin^4\theta_W\left(2-\frac{2E_R}{E_{\nu}}+\frac{E_R^2}{E_{\nu}^2}\right).
\end{equation}
Since $\sin^2\theta_W \approx 0.25$, the leading term in the $\nu_\mu e^-$ cross-section is suppressed, introducing a degeneracy between a measurement of $\sin^2\theta_W$ and the overall flux normalisation. However, the $\nu_e e^-$ channel does not suffer from this suppression and is highly sensitive to the value of $\sin^2\theta_W$. By providing high-intensity, pure beams of both $\nu_\mu$ and $\nu_e$ with precisely known fluxes, nuSTORM is uniquely capable of breaking this degeneracy and performing a high-precision measurement at low momentum transfer.

Our analysis follows the statistical framework highlighted in \secref{sec:statistical_framework} where the $\sin^2\theta_W$ is our model parameter and the SM prediction is calculated at the current global-fit value for low-$Q^2$ measurements, $\sin^2\theta_W = 0.23863$~\cite{ParticleDataGroup:2022pth,Alves:2024twb}. For nuSTORM most of the recoiled electrons will lie between $0.2\, \mathrm{GeV}<{E_R<6 \,\mathrm{GeV}}$ and hence the accessible $Q^2$ for nuSTORM would be in the range $(14-78 \, \mathrm{MeV})^2$.
The resulting sensitivity is shown in \figref{fig:chi_square_contour_weak} and compared with the projection for DUNE~\cite{deGouvea:2019wav}, SBND~\cite{Alves:2024twb}, and several other pre-existing measurements. Although DUNE will have a significantly larger exposure, nuSTORM's projected constraints are highly competitive. This excellent performance stems from its combination of a high-intensity flux, the availability of both $\nu_e$ and $\nu_\mu$ channels and robust control over systematic uncertainties. In particular, the clean $\nu_e$ beam from muon decays provides the necessary leverage to disentangle the effects of $\sin^2\theta_W$ from the flux normalisation. This enhanced sensitivity allows nuSTORM to constrain the weak mixing angle to $\sin^2\theta_W = 0.23863^{+0.00351}_{-0.00354}$, corresponding to a $1\sigma$ precision of about 1.5\%. For comparison, the optimistic SBND-PRISM benchmark at $10^{22}$ POT with $5\%$ correlated systematics yields a sensitivity of about 3\%~\cite{Alves:2024twb}. 
This result highlights nuSTORM's ability to conduct high-precision tests of the Standard Model that can help resolve existing tensions in electroweak observables.

\subsection{Neutrino Trident Production}\label{sec:trident}
%
\begin{figure}[t!]
    \centering
    \includegraphics[width=0.75\linewidth]{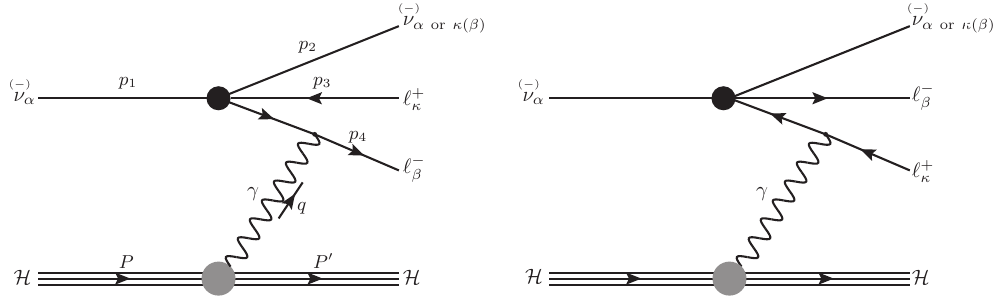}
    \caption{Feynman diagrams contributing to the neutrino trident process~\cite{Ballett:2018uuc}}
    \label{fig:Trident Feynamn diagram}
\end{figure}

nuSTORM's high-intensity beam and well-understood flux make it an ideal environment for studying rare Standard Model processes. One such process is neutrino trident production~(see \figref{fig:Trident Feynamn diagram}), where a neutrino scatters off a hadronic target $\mathcal{H}$, creating a pair of charged leptons~\cite{Czyz:1964zz,Lovseth:1971vv,Fujikawa:1971nx,Koike:1971tu}:

\begin{equation}
\overset{(-)}{\nu_\alpha} + \mathcal{H} \rightarrow \overset{(-)}{\nu}_{\alpha \, \text{or} \, \kappa(\beta) } \, + l_\beta^- + l_\kappa^+ +\mathcal{H}\,
\end{equation}
where $\{\alpha,\beta,\kappa\} \in \{e,\mu,\tau\}$. This process can be mediated by $W$ or $Z$ bosons. At the energy scales relevant for nuSTORM, the interaction is dominated by coherent scattering off the entire nucleus. The resulting lepton pair can be of the same or different flavours. To date, the only channel observed is dimuon production ($\nu_\mu+\mathcal{H} \rightarrow \nu_\mu + \mu^+ + \mu^- +\mathcal{H}$), first measured by the CHARM-II experiment~\cite{CHARM-II:1990dvf} and subsequently by CCFR~\cite{CCFR:1991lpl} and NuTeV~\cite{NuTeV:1998khj}.

\begin{figure}[t!]
    \centering
    \begin{subfigure}{0.49\linewidth}
        \centering 
        \includegraphics[width=1\textwidth]{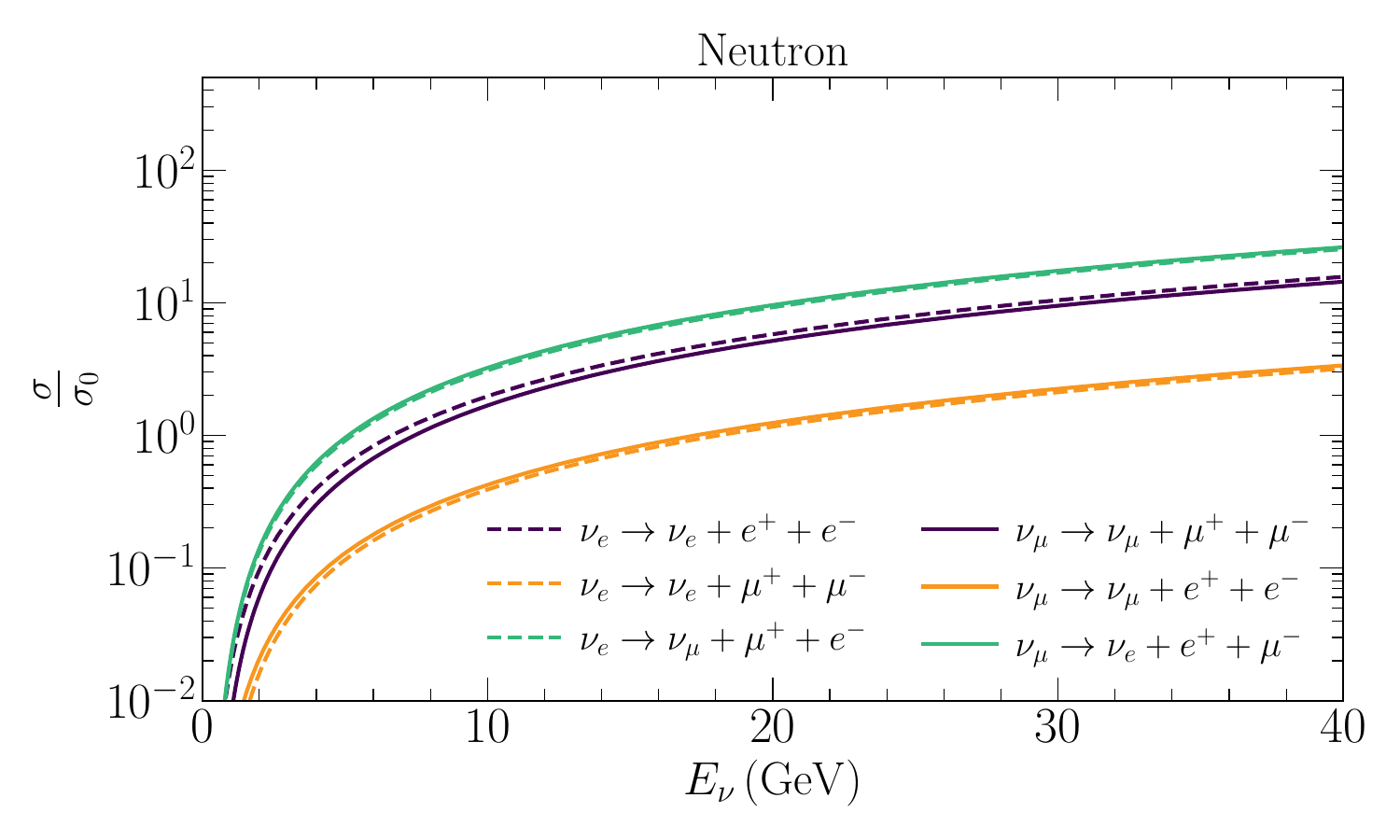}
        \caption{Cross Section for diffractive neutrino trident production on neutrons}
        \label{fig:diffractive Cross-section neutrons}
    \end{subfigure}
    \begin{subfigure}{0.49\linewidth}
        \centering 
        \includegraphics[width=1\textwidth]{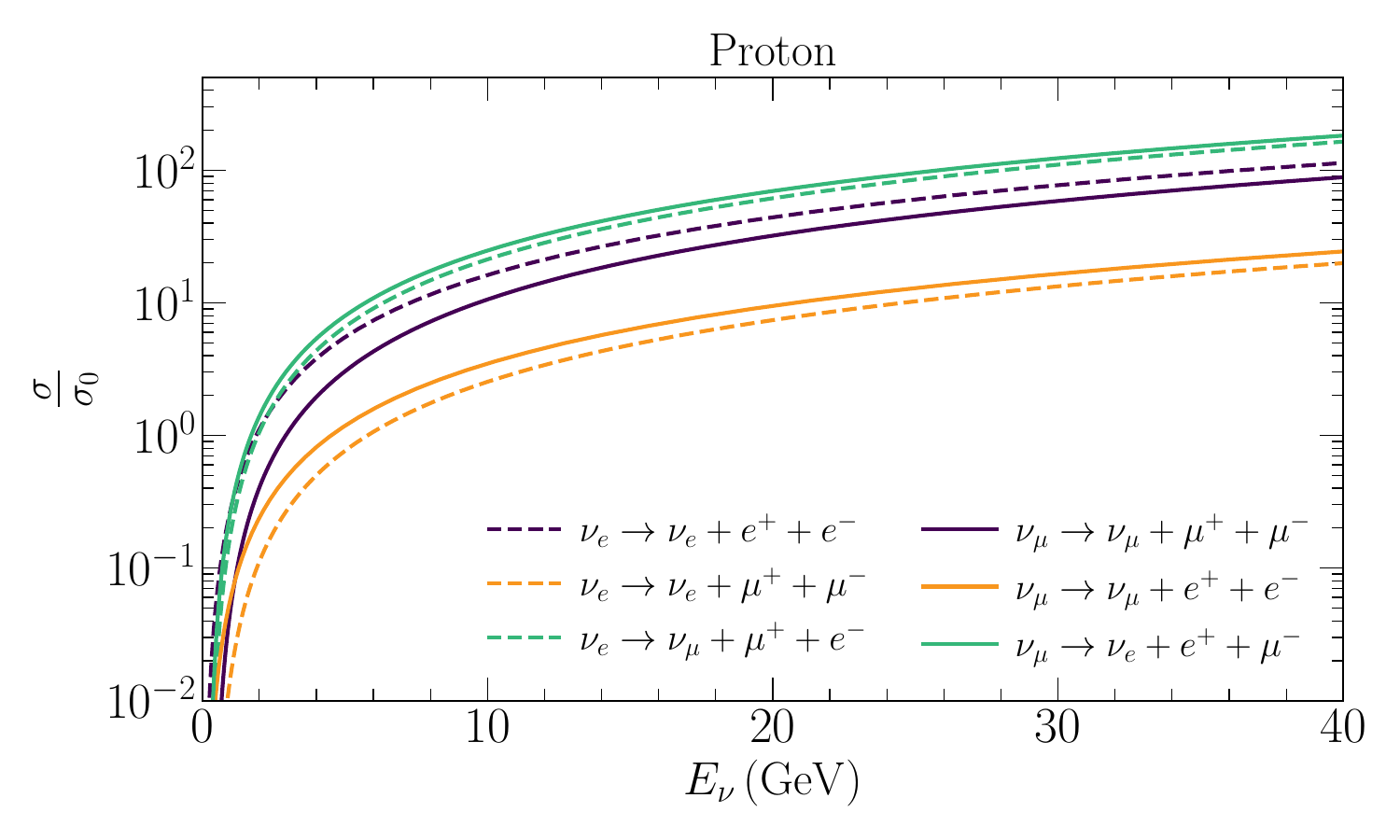}
        \caption{Cross-section for diffractive neutrino trident production on protons}
        \label{fig:diffractive Cross-section protons}
    \end{subfigure}\\[1ex]
    \begin{subfigure}{0.49\linewidth}
        \centering 
        \includegraphics[width=1.1\textwidth]{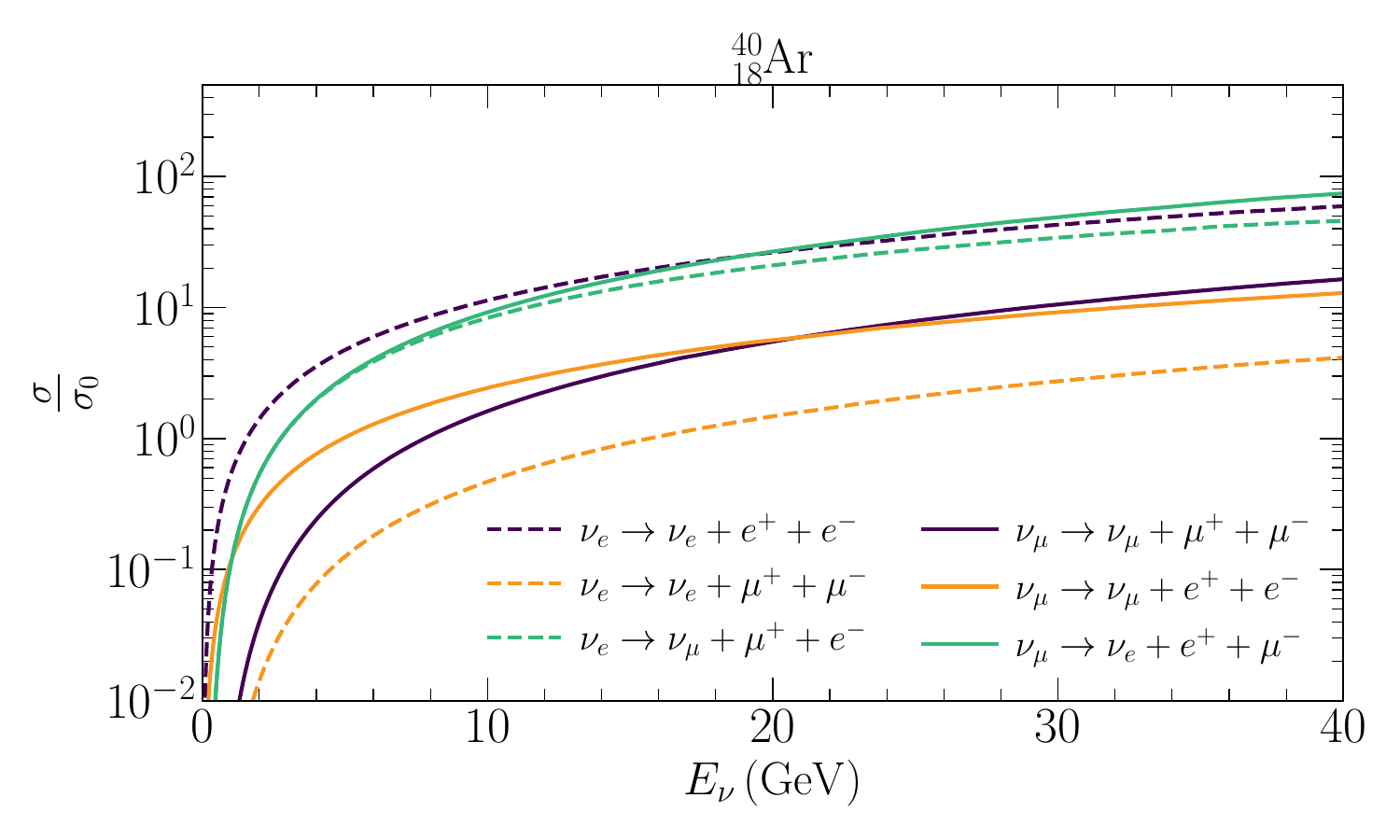}
        \caption{Cross-section for coherent neutrino trident production on $^{40}_{18}\text{Ar}$}
        \label{fig:coherent crosss section argon}
    \end{subfigure}
    \caption{Cross-section for diffractive (top row) and coherent (bottom row) trident production. The normalization $\sigma_0$ for the top row is $\sigma_0 = 10^{-44} \text{cm}^2$ and that for the bottom row is $\sigma_0 = \text{Z}^2\, 10^{-44} \text{cm}^2$ where Z here would be 18.} 
    \label{fig:trident_crosssecion}
\end{figure}
The trident cross-section has been computed in several theoretical approaches, including early four-fermion effective theories~\cite{Czyz:1964zz,Lovseth:1971vv,Fujikawa:1971nx}, intermediate-boson exchange models~\cite{Brown:1971} and the equivalent photon approximation (EPA)~\cite{Altmannshofer:2014,Magill:2017}. The EPA was widely used historically due to its simplicity, but it is now understood to be unreliable: while it can give a reasonable estimate for coherent dimuon production, it substantially overestimates the cross-section for channels with final-state electrons or in the diffractive regime, in some cases by more than 200\%~\cite{Ballett:2018uuc}. For this reason, we employ the complete $2 \rightarrow 4$ calculation presented in Ref.~\cite{Ballett:2018uuc}, which provides a robust description across all channels and targets.

\begin{figure}[t]
    \centering
    \begin{subfigure}[b]{0.49\textwidth}
        \centering 
        \includegraphics[width=1\textwidth]{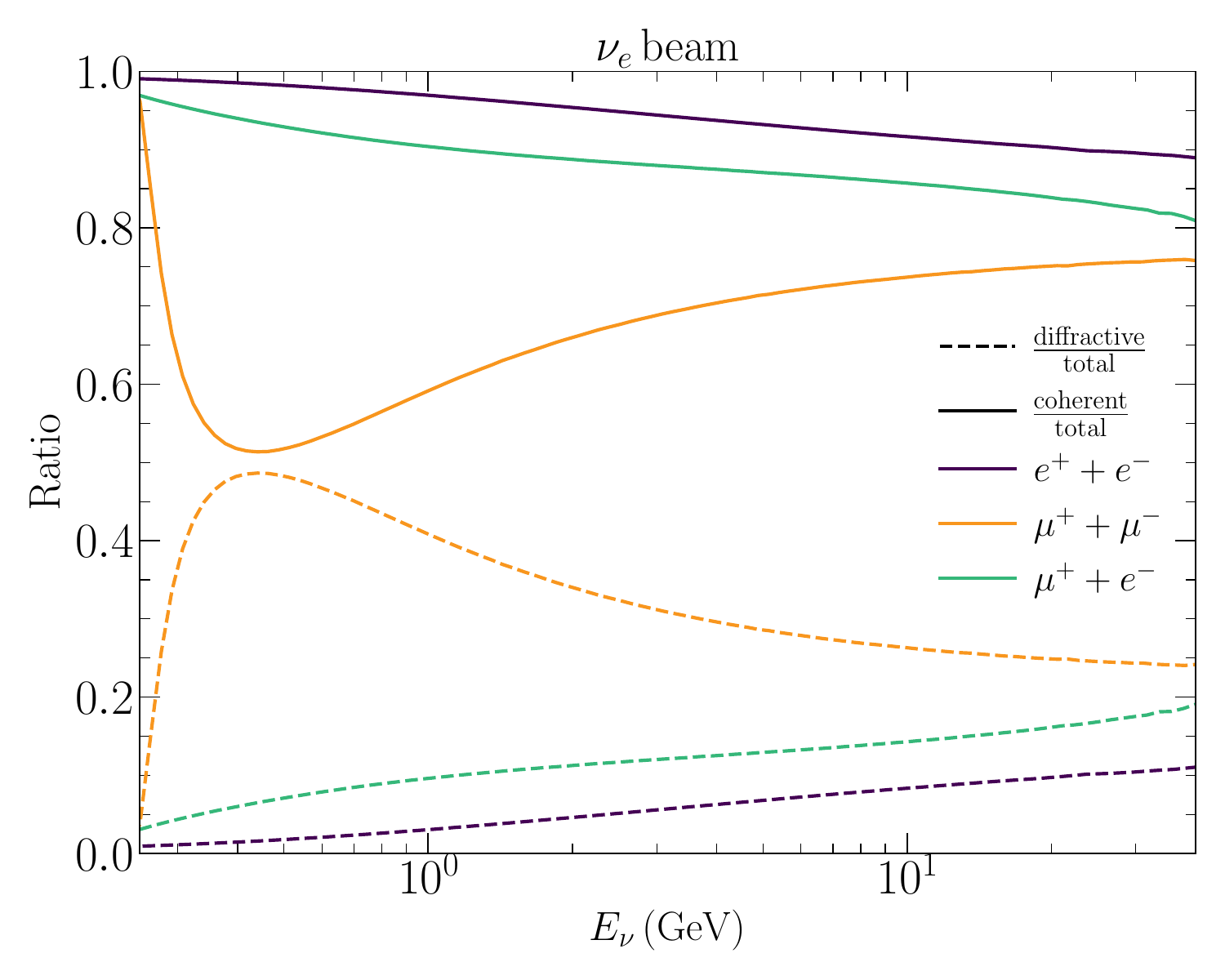}
        \label{fig:diffractive_coherent_ratio_nu_e}
    \end{subfigure}
    \begin{subfigure}[b]{0.49\textwidth}
        \centering 
        \includegraphics[width=1\textwidth]{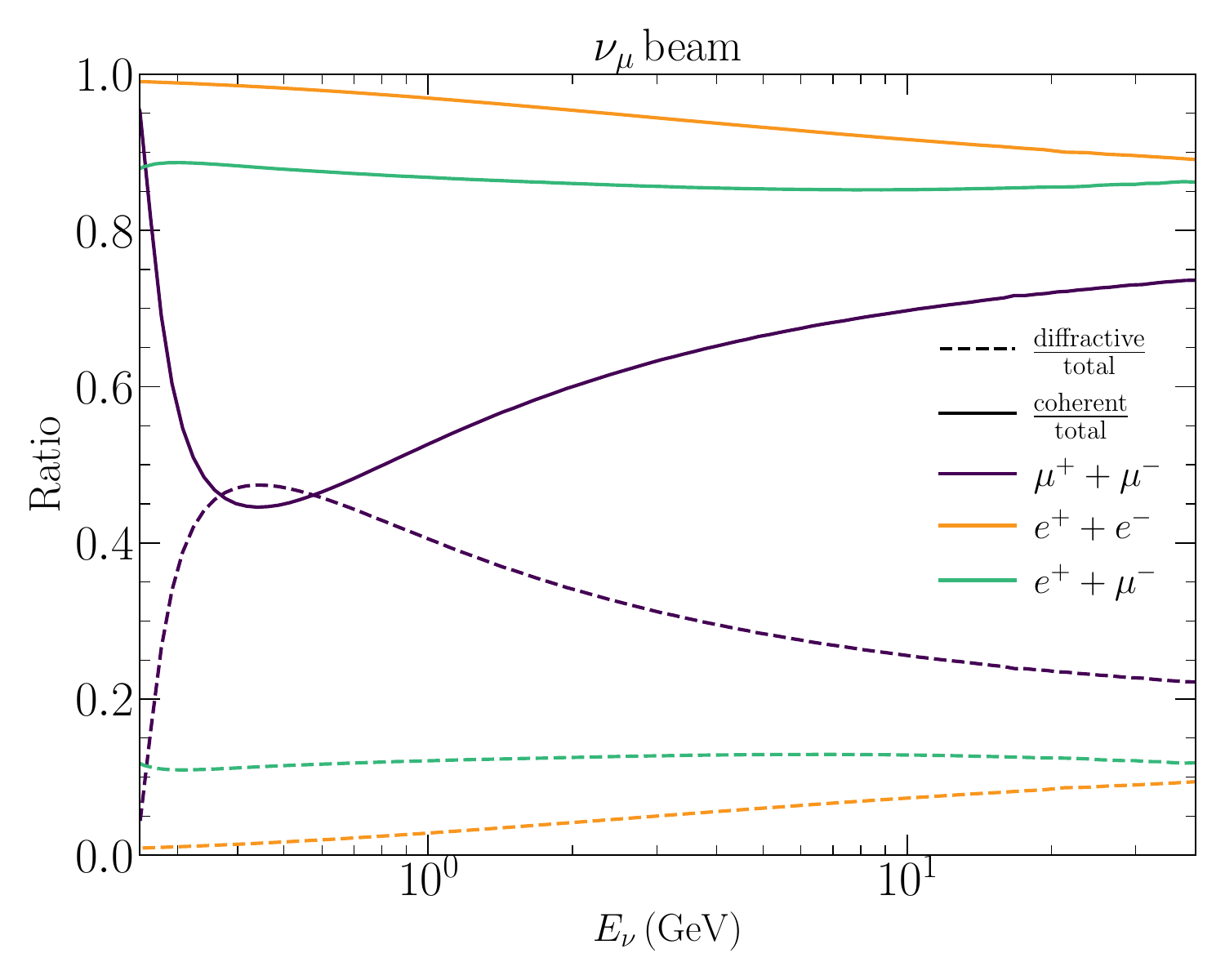}
        \label{fig:diffractive_coherent_ratio_nu_mu}
    \end{subfigure}
    \caption{Ratio of the coherent (solid) and diffractive (dashed) contributions to the total trident cross-section for $\nu_e$ (left) and $\nu_\mu$ (right) beams. The legend indicates the final-state channels: purple corresponds to processes with both CC and NC contributions, orange to purely NC channels and green to purely CC channels.}
    \label{fig:coherent_diffractive_ratio}
\end{figure} 

The trident cross-section as a function of neutrino energy is shown in Fig.~\ref{fig:trident_crosssecion} for several flavour combinations. The top two plots show the cross-section where the hadronic target is the neutron and proton, while the bottom plot shows the cross-section on Argon. 
\figref{fig:coherent_diffractive_ratio} shows the relative contributions of the coherent and diffractive regimes to the total trident cross-section. For channels with electrons in the final state, the coherent component dominates. This is because electron production requires only a small momentum transfer, favouring interactions in which the neutrino couples coherently to the entire nucleus. By contrast, in channels with two muons in the final state, the larger mass threshold demands a higher momentum transfer and as a result the diffractive contribution becomes comparable to the coherent one. 
A further feature is that in the dimuon channel for $\nu_\mu$ the diffractive and coherent contributions exhibit a crossover, whereas no such behaviour occurs for $\nu_e$ induced dimuon production. This difference arises from the underlying electroweak structure: $\nu_\mu$ tridents receive contributions from both CC and NC interactions, enhancing the overall cross-section, while $\nu_e$ tridents proceed only through NC. 

Experimentally, coherent and diffractive tridents yield identical charged-lepton final states, but they may be distinguished statistically through their hadronic signatures and momentum transfer distributions. Coherent events are characterised by very low hadronic recoil, while diffractive interactions are accompanied by additional nucleons or nuclear breakup products and extend to larger $Q^2$. Detectors with sensitivity to low-energy hadronic activity, such as LArTPCs, therefore offer the possibility of partially separating the two contributions. The precisely characterised flux at nuSTORM would provide an ideal testbed for such an analysis.
\begin{table}[t!]
    \centering
    \renewcommand{\arraystretch}{1.2} 
    \begin{tabular}{lccccc}
        \toprule
        \textbf{Channel} & \textbf{SBND} & $\bm{\mu}$\textbf{BooNE} & \textbf{ICARUS} & \textbf{DUNE} & \textbf{nuSTORM} \\
        \midrule
        \multirow{2}{*}{$e^{\pm}\mu^{\mp}$} 
            & 10   & 0.7 & 1.0 & 2993 (2307) & 173 \\
            & 2.0  & 0.1 & 0.2 & 692 (530)   & 29  \\
        \midrule
        \multirow{2}{*}{$e^{+}e^{-}$} 
            & 6.0  & 0.4 & 0.7 & 1007 (800)  & 107 \\
            & 0.7  & 0.0 & 0.1 & 143 (111)   & 5   \\
        \midrule
        \multirow{2}{*}{$\mu^{+}\mu^{-}$} 
            & 0.4  & 0.0 & 0.0 & 286 (210)   & 14  \\
            & 0.4  & 0.0 & 0.0 & 196 (147)   & 9   \\
        \bottomrule
    \end{tabular}
    \caption{Comparison of the total number of trident events across different experiments~\cite{Ballett:2018uuc}. 
    For each channel, the top row corresponds to \emph{coherent} production, while the bottom row corresponds to \emph{diffractive} production. 
    For DUNE, the numbers in parentheses indicate the antineutrino running mode.}
    \label{Table: Trident Number of events}
\end{table}

At nuSTORM, the neutrino flux from pion decays is nearly two orders of magnitude larger than that from kaon decays (see Fig.~12 of Ref.~\cite{Adey:2015iha}). Moreover, undecayed kaons are absorbed at the end of the production straight, further suppressing their contribution. We, therefore, restrict our analysis to neutrinos originating from pion decays. This choice is well-justified, as it captures the dominant flux component while providing a conservative framework for the study of neutrino trident production. 

The event rate is calculated by employing the methodology outlined in~\secref{sec:event_estimation}. Table~\ref{Table: Trident Number of events} compares the projected event rates at nuSTORM with other experiments. The results show that nuSTORM's reach is highly competitive, surpassed only by DUNE, which benefits from a higher exposure. Potential backgrounds from rare meson resonances or decays like $K^{ \pm} \rightarrow \mu^+ + \mu^- +\pi^{ \pm}$ via BSM mediators are rare and expected to be distinguishable due to different kinematics or energy ranges. The primary remaining background is therefore expected to arise from particle misidentification. An extensive study of such backgrounds has been conducted in~\cite{Ballett:2018uuc}. 

\section{Searches for Physics Beyond the Standard Model at nuSTORM}
In addition to its programme of precision Standard Model measurements, nuSTORM serves as a powerful probe for new physics. This section details the facility's sensitivity to several beyond the Standard Model scenarios, including searches for sterile neutrinos, large extra dimensions, lepton flavour violation and axion-like particles.
\subsection{Sterile Neutrinos}\label{sec:steriles}
Although neutrino oscillation parameters have been measured with increasing precision using reactor, solar, atmospheric and accelerator data~\cite{KamLAND:2004mhv,SNO:2002tuh,Super-Kamiokande:1998kpq,MINOS:2008kxu,DayaBay:2012fng,DoubleChooz:2011ymz,RENO:2012mkc}, the standard three-neutrino paradigm cannot explain several longstanding experimental anomalies. A simple extension to the Standard Model that could account for these is the introduction of additional light, non-interacting ``sterile'' neutrinos that mix with the active flavours.

A persistent hint for such states comes from the LSND~\cite{LSND:2001aii} and MiniBooNE~\cite{MiniBooNE:2018esg} experiments, which reported significant excesses of low-energy electron-like events. While these excesses could be interpreted as $\nu_\mu \to \nu_e$ oscillations in a (3+1) model with $\Delta m^2 \sim 1\,\mathrm{eV}^2$, this explanation is in tension with data from disappearance experiments~\cite{Kopp:2013vaa,Dentler:2018sju}. The MicroBooNE experiment, using high-resolution Liquid Argon Time Projection Chambers (LArTPC), found no evidence of an electron-neutrino excess in multiple analyses~\cite{MicroBooNE:2021zai,MicroBooNE:2021tya,MicroBooNE:2021nxr}. These results strongly disfavour several interpretations of the MiniBooNE result, though the full scope of this conclusion remains a subject of debate in the literature~\cite{Arguelles:2021meu}.

Further anomalies suggesting eV-scale oscillations have been reported in the reactor and gallium experiments. The reactor antineutrino anomaly refers to a deficit in the observed $\bar\nu_e$ flux compared to theoretical predictions~\cite{Mention:2011rk}, while the gallium anomaly describes a similar deficit seen in the calibration of the GALLEX and SAGE solar neutrino experiments~\cite{Giunti:2019aiy}. However, modern very-short-baseline reactor experiments (such as DANSS~\cite{DANSS:2018fnn}, NEOS~\cite{NEOS:2016wee}, STEREO~\cite{STEREO:2018blj} and PROSPECT~\cite{PROSPECT:2018dtt}) have placed stringent constraints on the parameter space favoured by these anomalies, significantly reducing their statistical significance in global fits~\cite{DANSS:2018fnn,NEOS:2016wee,PROSPECT:2020sxr}. Global analyses find that while hints of new physics persist, there is strong tension between appearance and disappearance datasets, with no single parameter region providing a consistent explanation~\cite{Dentler:2018sju,Berryman:2021yan,Kopp:2013vaa}. These anomalies are typically studied in a minimal (3+1) scenario, parameterised by the effective mixing angles $\sin^2 2\theta_{ee}$, $\sin^2 2\theta_{\mu\mu}$ and $\sin^2 2\theta_{e\mu}$.

Sterile neutrinos in this mass range also have significant cosmological implications, as they would alter the effective number of neutrino species ($N_{\rm eff}$)~\cite{Abazajian:2012ys, Berryman:2019nvr} and contribute to the sum of neutrino masses. Models with eV-scale sterile neutrinos must satisfy the stringent cosmological bounds obtained within $\Lambda$CDM~\cite{Archidiacono:2012ri}, or else invoke new physics (such as dynamical dark energy) that modifies the standard cosmological history. Dynamical dark energy extensions (e.g.\ $w_0w_a$CDM+sterile) can yield an improved fit relative to $\Lambda$CDM+sterile~\cite{Du:2025iow,Pan:2023frx}, and the recent DESI results~\cite{DESI:2025zgx} further motivate considering such extensions.
The nuSTORM facility is poised to decisively test these scenarios. Its stored-muon beam provides a precisely characterised neutrino source, enabling powerful probes of both appearance and disappearance channels with percent-level systematic uncertainties~\cite{nuSTORM:2013cqr}.
In the minimal $(3+1)$ scenario, the $\nu_\alpha \to \nu_\beta$ transition probability in the short-baseline (SBL) limit is~\cite{Penedo:2022etl,Kopp:2013vaa}
\begin{equation}
P(\nu_{\alpha} \rightarrow \nu_{\beta}) = \delta_{\alpha \beta}
- \sin^2 2\theta_{\alpha \beta} \, \sin^2\!\left(\frac{\Delta m^2_{41}L}{4E}\right)\,,
\end{equation}
where the effective mixing is defined as $\sin^2 2\theta_{\alpha\beta} \equiv 4|U_{\alpha4}|^2 |U_{\beta4}|^2$ and $\Delta m^2_{41} = m^2_4 - m^2_1$. The disappearance and appearance channels are related by
\begin{equation}
\sin^2 2\theta_{e\mu}
= 4\,|U_{e4}|^2 |U_{\mu4}|^2
\simeq \tfrac14\,\sin^2 2\theta_{ee}\,\sin^2 2\theta_{\mu\mu}\,.
\end{equation}
\begin{figure}[t!]
    \centering
    \begin{subfigure}[b]{0.49\textwidth}
        \centering 
        \includegraphics[width=0.99\linewidth]{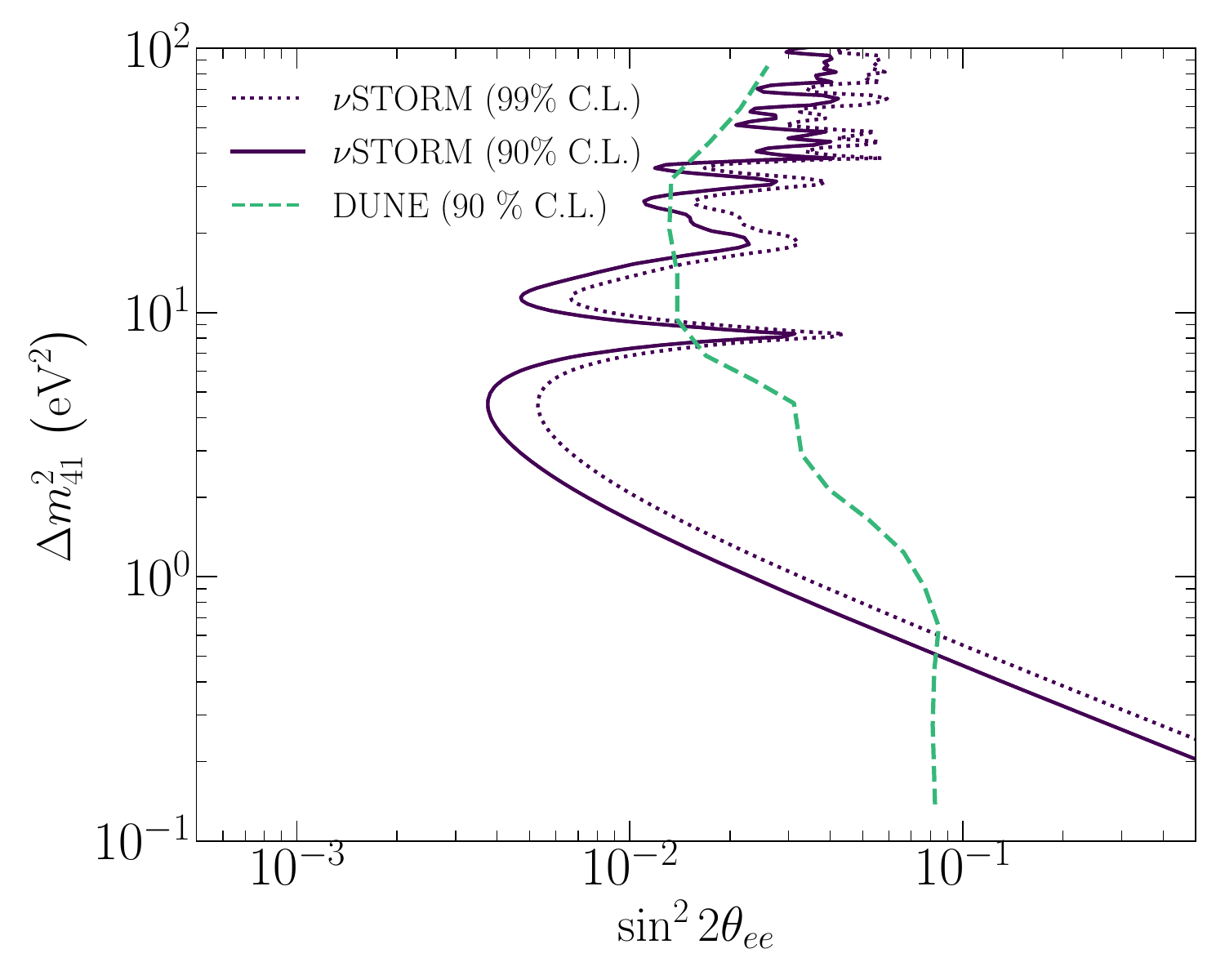}
        \label{fig:chi_square_sterile_nu_e}
    \end{subfigure}
    \begin{subfigure}[b]{0.49\textwidth}
         \centering
        \includegraphics[width=0.99\linewidth]{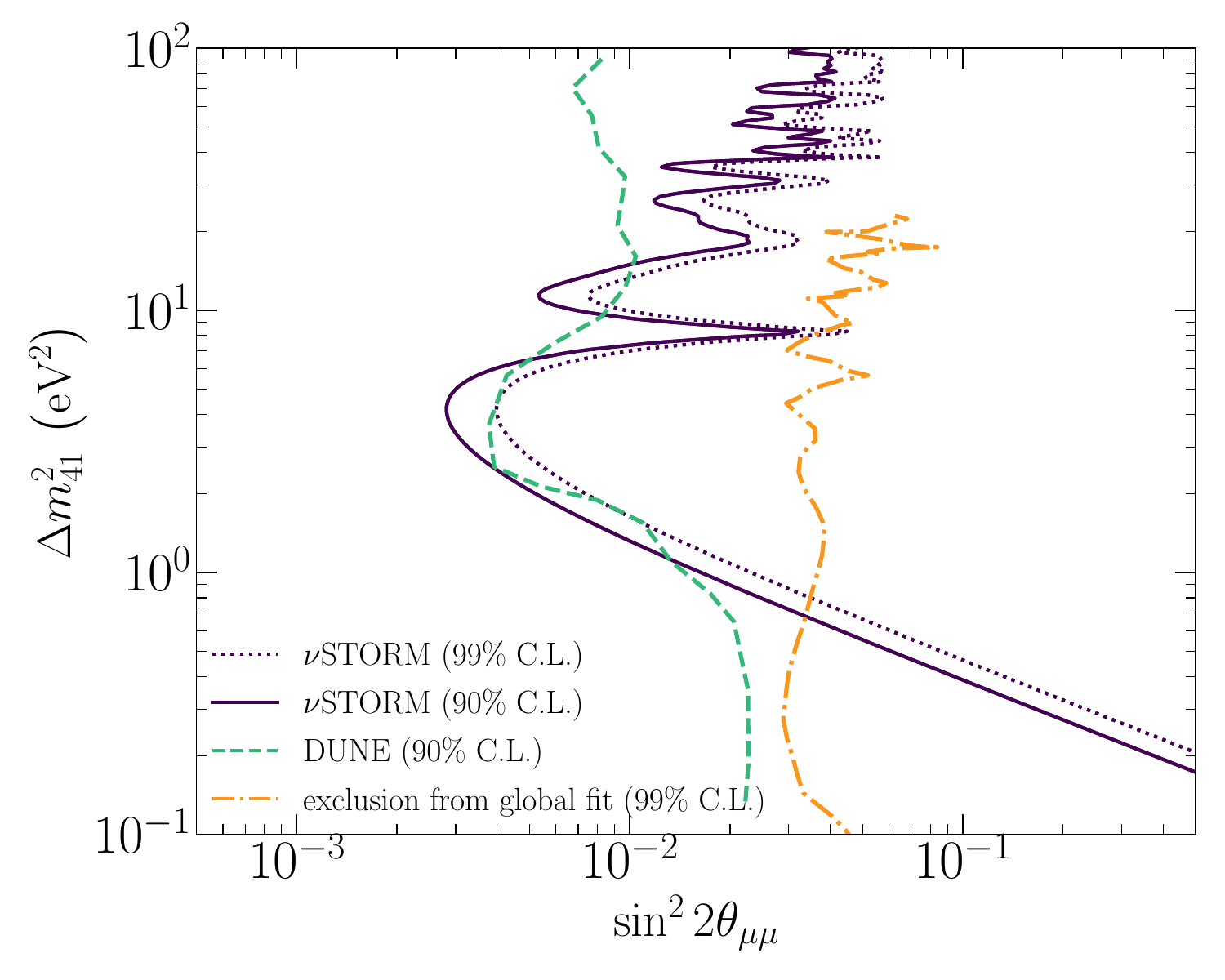}
        \label{fig:chi_square_sterile_nu_mu}
    \end{subfigure}
    \caption{Sensitivity to $\nu_e \rightarrow \nu_e$ (left) and $\overset{(-)}{\nu}_{\mu} \rightarrow \overset{(-)}{\nu}_{\mu}$ (right) disappearance under the assumption of 1\% systematic uncertainty. The exclusion from global fit is obtained from ~\cite{Dentler:2018sju} and is used for the $\nu_\mu$ channel. Results for DUNE analysis are obtained for \cite{Penedo:2022etl}} 
    \label{fig:chi_square_sterile neutrino}
\end{figure}

This relationship means that strong constraints from disappearance channels tightly restrict the allowed appearance signal. Global fits~\cite{Dentler:2018sju,Berryman:2021yan,Kopp:2013vaa} indicate that much of the parameter space suggested by appearance anomalies is excluded by disappearance searches. nuSTORM, with its precisely known flux and robust control of systematics, is uniquely positioned to test the remaining parameter space and deliver world-leading sensitivity to the sterile neutrino hypothesis.

To constrain the sterile neutrino parameter space, we perform a $\chi^2$ analysis of the neutrino survival probability. We account for smearing effects from the extended muon decay region by assuming a uniform neutrino production baseline between $50 < L\, (\mathrm{m}) < 250$. One could, in principle, exploit the variable momentum operation of nuSTORM to enhance sensitivity to different values of $\Delta m_{41}^2$. However, this effect is already effectively incorporated through baseline smearing and the substantial statistics at lower energy modes, which extend the accessible $\Delta m_{41}^2$ range without requiring dedicated momentum scans.

We make use of the $\chi^2$ analysis outlined in~\secref{sec:statistical_framework} with $\vec\theta = (\Delta m_{41}^2,\sin^2\theta_{\alpha \alpha})$ as our model parameters. For the standard oscillation parameters, we use the latest best-fit values from global analyses~\cite{Esteban:2024eli}.
\figref{fig:chi_square_sterile neutrino} presents our projected sensitivity in the $(\Delta m_{41}^2,\sin^2\theta_{\alpha \alpha})$ plane. In the $\bar{\nu}_{\mu}$ disappearance channel, nuSTORM's sensitivity significantly surpasses existing experimental limits, primarily due to its intense muon neutrino beam from both muon and pion decays. It is also competitive to DUNE's projected sensitivity in the region $\Delta m_{41}^2\simeq \mathcal{O}(1-10)\, \mathrm{eV}^2$. Similarly, in the $\nu_e$ disappearance channel, nuSTORM's access to a pure, high-intensity electron neutrino beam allows it to set stronger limits than DUNE for $\Delta m_{41}^2~(\mathrm{eV}^2) \lesssim 10$.

\subsection{Large Extra Dimensions}\label{sec:LED}
Large Extra Dimensions (LED) embed the familiar \(3+1\) spacetime into a higher-dimensional bulk where only gauge singlets, such as right-handed neutrinos, may propagate~\cite{Arkani-Hamed:1998wuz,Antoniadis:1998ig}. Adopting the flat-metric framework of Refs.~\cite{Mohapatra:1999zd,Mohapatra:2000wn,Machado:2011jt}, we assume a single large extra dimension of radius \(R_{\mathrm{LED}}\), giving an effectively five-dimensional theory. Compactification imposes periodicity on the right-handed neutrino fields, $\nu_{R}$, which produces a Kaluza–Klein (KK) tower of four-dimensional fields \(\nu_{R}^{(N)}\) with masses \(N/R_{\mathrm{LED}}\) for $N=1,2,\dots$. Mixing between this tower and the brane’s left-handed neutrinos distorts the oscillation pattern, providing a direct probe of the extra dimension's radius, \(R_{\mathrm{LED}}\).
The four-dimensional mass term that governs neutrino oscillations in the LED framework is~\cite{Machado:2011jt}:
\begin{equation}
-\mathcal{L}\supset \sum_{\alpha, \beta} m_{\alpha \beta}^D\left[\overline{\nu}_{\alpha L}^{(0)} \nu_{\beta R}^{(0)}+\sqrt{2} \sum_{N=1}^{\infty} \overline{\nu}_{\alpha L}^{(0)} \nu_{\beta R}^{(N)}\right]+\sum_\alpha \sum_{N=1}^{\infty} \frac{N}{R_{LED}} \overline{\nu}_{\alpha L}^{(N)} \nu_{\alpha R}^{(N)}+\text{h.c.}\,,
\end{equation}
where $\alpha,\beta = e,\mu,\tau$ are the three leptonic flavours, \(\nu^{(0)}_{\alpha L}\) are the usual SM (brane) neutrinos, \(\nu^{(0)}_{\alpha R}\) are the zero-mode bulk partners and \(\nu^{(N)}_{\alpha R,L}\;(N\ge1)\) are the \(N\)-th KK excitations. The factor \(\sqrt{2}\) arises from the normalisation of the higher KK wavefunctions. The Dirac matrix \(m^{D}_{\alpha\beta}\) can be diagonalised by biunitary rotations, \(U^{\dagger} m^{D} R = \mathrm{diag}(m^{D}_{1},m^{D}_{2},m^{D}_{3})\), yielding the diagonal elements $m_i^D$. After this step, the three flavour sectors decouple and for each index \(i=1,2,3\), the mass term is
\begin{equation}\label{eq:LED_block}
\mathcal{L}\supset -\overline{\nu_{iL}}\begin{pmatrix}
m_i^D & \sqrt{2}m_i^D &\sqrt{2}m_i^D &\dots& \sqrt{2}m_i^D\\
0 & \mu_1 &0&\dots & 0\\
0 & 0 & \mu_2 &\dots & 0\\
\dots&\dots&\dots&\dots&\dots\\
0 & 0 & 0 &\dots & \mu_N
\end{pmatrix}\nu_{iR}\equiv -\overline{\nu_{iL}}\mathbf{M}\nu_{iR}\,,
\end{equation}

where \(\nu_{iL}\!=\!(\nu^{(0)}_{iL},\nu^{(1)}_{iL},\nu^{(2)}_{iL},\dots)^{T}\) and likewise for \(\nu_{iR}\). The first row contains the Dirac mass that couples the active state to every KK mode, while the remaining rows determine the KK masses, with \(\mu_N=N/R_{\mathrm{LED}}\). Diagonalising this infinite matrix yields the mass eigenvalues and mixings used in the subsequent oscillation analysis.

Working in the basis where the {Dirac} matrix, \(m^{D}\), is already diagonal, the flavour states are 
\be
\nu_{\alpha L}^{(0)}=\sum_{i}U_{\alpha i}\,\nu_{iL}^{(0)}\,,\qquad
\nu_{\alpha R}^{(0)}=\sum_{i}R_{\alpha i}\,\nu_{iR}^{(0)}\,,\qquad
\nu_{\alpha L,R}^{(N)}=\sum_{i}R_{\alpha i}\,\nu_{iL,R}^{(N)}\, ,
\ee
with \(U\) being the usual PMNS matrix and \(R\) the rotation acting on the right-handed sector. For each generation, \(i=1,2,3\), the infinite KK mass matrix from Eq.~\eqref{eq:LED_block} is denoted \(\boldsymbol{M}_{i}\); its eigenvalues are found from \(\boldsymbol{M}_{i}^{\dagger}\boldsymbol{M}_{i}\). The condition \(\det(\boldsymbol{M}_{i}^{\dagger}\boldsymbol{M}_{i}-m^{2}\mathbb{1})=0\) gives the following transcendental relation\,\cite{Machado:2011jt,Siyeon:2024pte}
\begin{equation}
\label{eq:LED_eig}
\frac{m_{i}^{\,n}}{R_{\mathrm{LED}}} = \pi\,(m^{D}_{i})^{2}\cot\bigl(\pi\,m_{i}^{\,n}R_{\mathrm{LED}}\bigr),
\qquad n = 0,1,2,\dots .
\end{equation}
The roots of this equation, \(m_{i}^{\,n}\), are the physical masses. The $n=0$ modes reproduce the three light neutrinos, where $m_{1}^{(0)}$ is allowed to float and the remaining masses are fixed by the measured squared differences \(\Delta m^{2}_{21}=m_{2}^{(0)2}-m_{1}^{(0)2}\) and \(\Delta m^{2}_{31}=m_{3}^{(0)2}-m_{1}^{(0)2}\). The coupling of each eigenstate to the weak gauge bosons is weighted by
\be
\label{eq:overlap_rewrite}
|V_{0i}^{\,n}|^{2}=
\frac{2}{1+\bigl(m_{i}^{\,n}/m^{D}_{i}\bigr)^{2}
+ (m^{D}_{i}R_{\mathrm{LED}})^{2}} \,.
\ee
A neutrino created in a charged-current interaction will be the linear combination of all possible mass eigenstates
\be
\label{eq:flavour_state}
\nu_{\alpha}^{(0)}=
\sum_{i=1}^{3}\sum_{n=0}^{\infty}
U_{\alpha i}\,V_{0i}^{\,n}\,\nu_{i}^{\,n}\,.
\ee
From this, the survival probability of a flavour eigenstate neutrino propagating over a distance \(L\) with energy \(E\) is
\begin{equation}
P(\nu_{\alpha}^{(0)} \rightarrow \nu_{\alpha}^{(0)}) = \left|\sum_i^3 \sum_n^\infty |U_{\alpha i}|^2 (V_{0i}^n)^2 \exp \left(-i\frac{\left(m_i^n\right)^2L}{2E}\right)\right|^2\,.
\end{equation}
In the limit $R_{\mathrm{LED}}\ll 1$, the KK eigenvalue equation implies that the lightest mode satisfies $m_i^{0}\to m_i^{D}$, while higher modes scale as $m_i^{n>0}\simeq n/R_{\mathrm{LED}}\to\infty$ and decouple; in this regime, the oscillations reduce to the standard three-flavour form.
\begin{figure}[t!]
    \centering
    \begin{subfigure}[b]{0.49\textwidth}
        \centering 
        \includegraphics[width=1.1\textwidth]{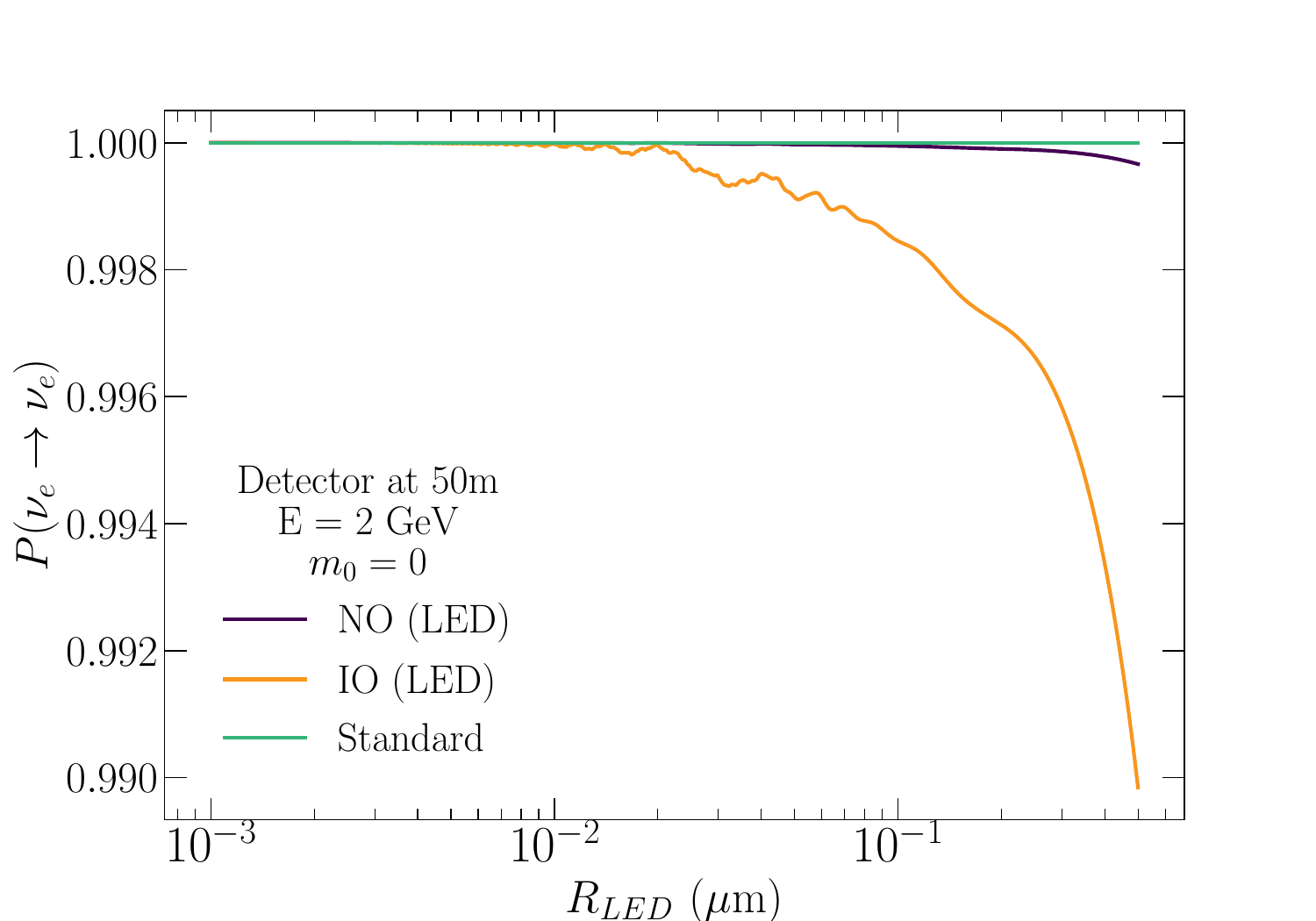}
        \label{fig:led_oscillation_vs_energy_nue}
    \end{subfigure}
    \begin{subfigure}[b]{0.49\textwidth}
        \centering 
        \includegraphics[width=1.1\textwidth]{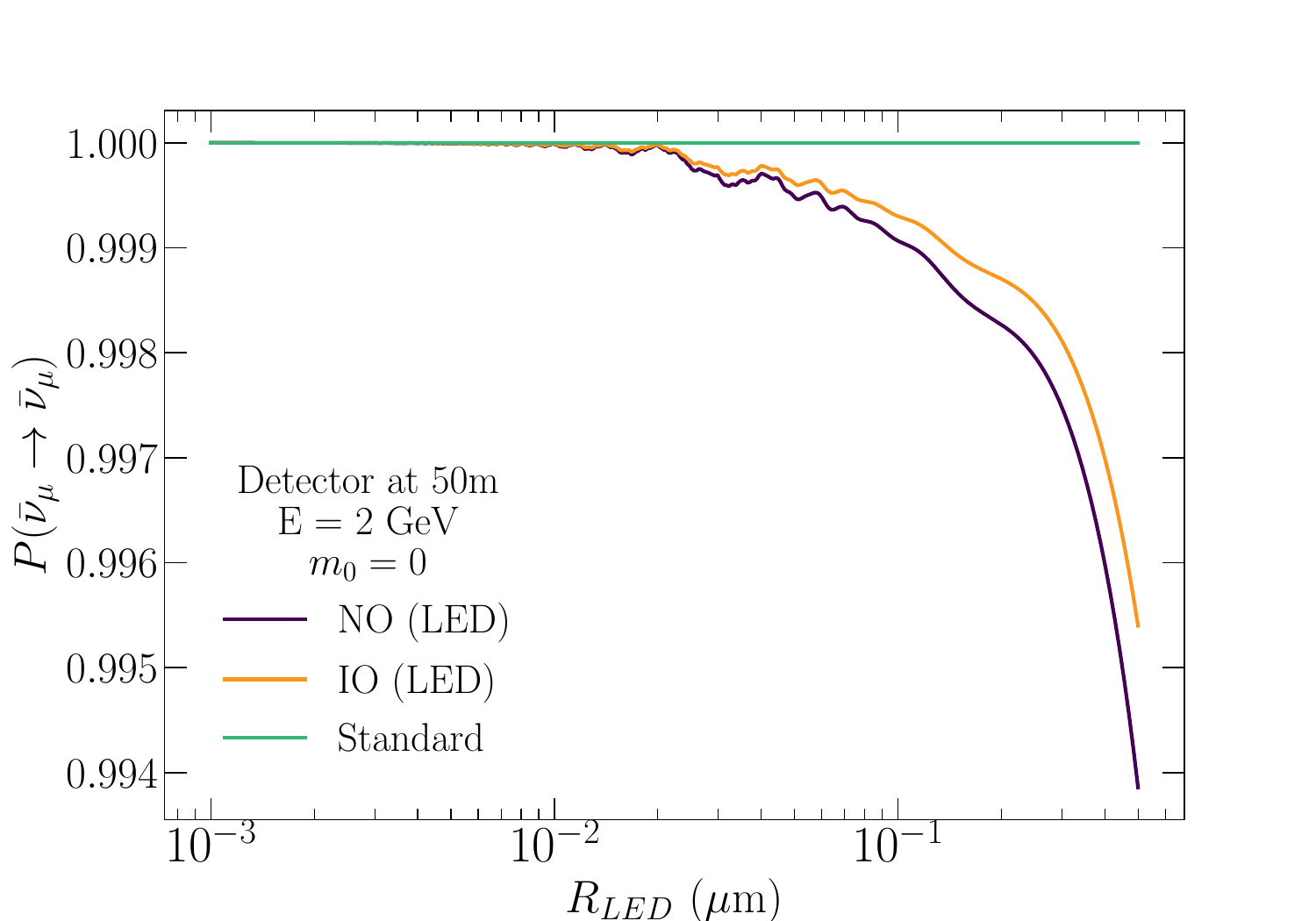}
        \label{fig:led_oscillation_vs_energy_numu}
    \end{subfigure}
    \caption{Impact of Large Extra Dimensions (LED) on the short baseline disappearance of $\nu_e$ and $\nu_\mu$ neutrinos. The plot illustrates how LED modifies the disappearance probability in the two channels compared to the standard three flavour scenario.
    }
    \label{fig1:led_oscillation_vs_energy}
\end{figure} 
For $\nu_\mu\to\nu_\mu$ disappearance, the relevant contributions to the survival probability are weighted by terms of the form $|U_{\mu i}|^2 (V_{0i}^n)^2$. Since the $|U_{\mu i}|^2$ elements are all of comparable magnitude, while $(V_{0i}^n)^2\propto (m_i^D)^2$ in the limit $m_i^DR_{\mathrm{LED}}\ll1$~\cite{Machado:2011jt,Siyeon:2024pte}, the contributions are not dominated by a particular mass eigenstate. Consequently, the relative size of LED effects in the $\nu_\mu$ channel is only weakly sensitive to the mass ordering, showing a slight enhancement for Normal Ordering. 

By contrast, for $\nu_e\to\nu_e$ disappearance the weights $|U_{e i}|^2 (V_{0i}^n)^2$ are dominated by the $i=1,2$ components because $|U_{e3}|\ll |U_{e1}|,|U_{e2}|$. In the normal ordering (NO), where $m_3\gg m_{1,2}$, any large $(V_{03}^n)^2\propto (m_3^D)^2$ contribution is suppressed by the small $|U_{e3}|^2$. In the inverted ordering (IO), however, the heavier states correspond to larger weights and hence, LED signatures are expected to be more prominent in the electron channel. The oscillation probabilities shown in \figref{fig1:led_oscillation_vs_energy} illustrate this behaviour, where the LED modification to the $\nu_{\mu}$ channel is similar for both orderings, whereas the modification for the $\nu_e$ channel is highly suppressed for NO and significant only for IO. 

To explore deviations from the standard three-neutrino paradigm in the presence of Large Extra Dimensions, we scan over the relevant parameter space using a $\chi^2$ statistic similar to the sterile neutrino case. Two of the three Dirac mass parameters, $m_i^D$, are fixed using the measured mass-squared differences and we take the lightest Dirac mass as a free parameter, denoted $m_0$ (where $m_0 \equiv m_1^D$ for Normal Ordering and $m_0 \equiv m_3^D$ for Inverted Ordering). The remaining free parameter is the radius of the extra dimension, $R_{\mathrm{LED}}$.
The statistical comparison between the LED hypothesis and the standard oscillation model is performed using the $\chi^2$ function from~\secref{sec:statistical_framework}. Here the model parameters are defined as $\vec\theta = (m_0,R_{LED})$ and the standard oscillation parameters are obtained from global fit data~\cite{Esteban:2024eli}.

\begin{figure}[t!]
    \centering
    \begin{subfigure}[b]{0.49\textwidth}
        \centering 
        \includegraphics[width=1\linewidth]{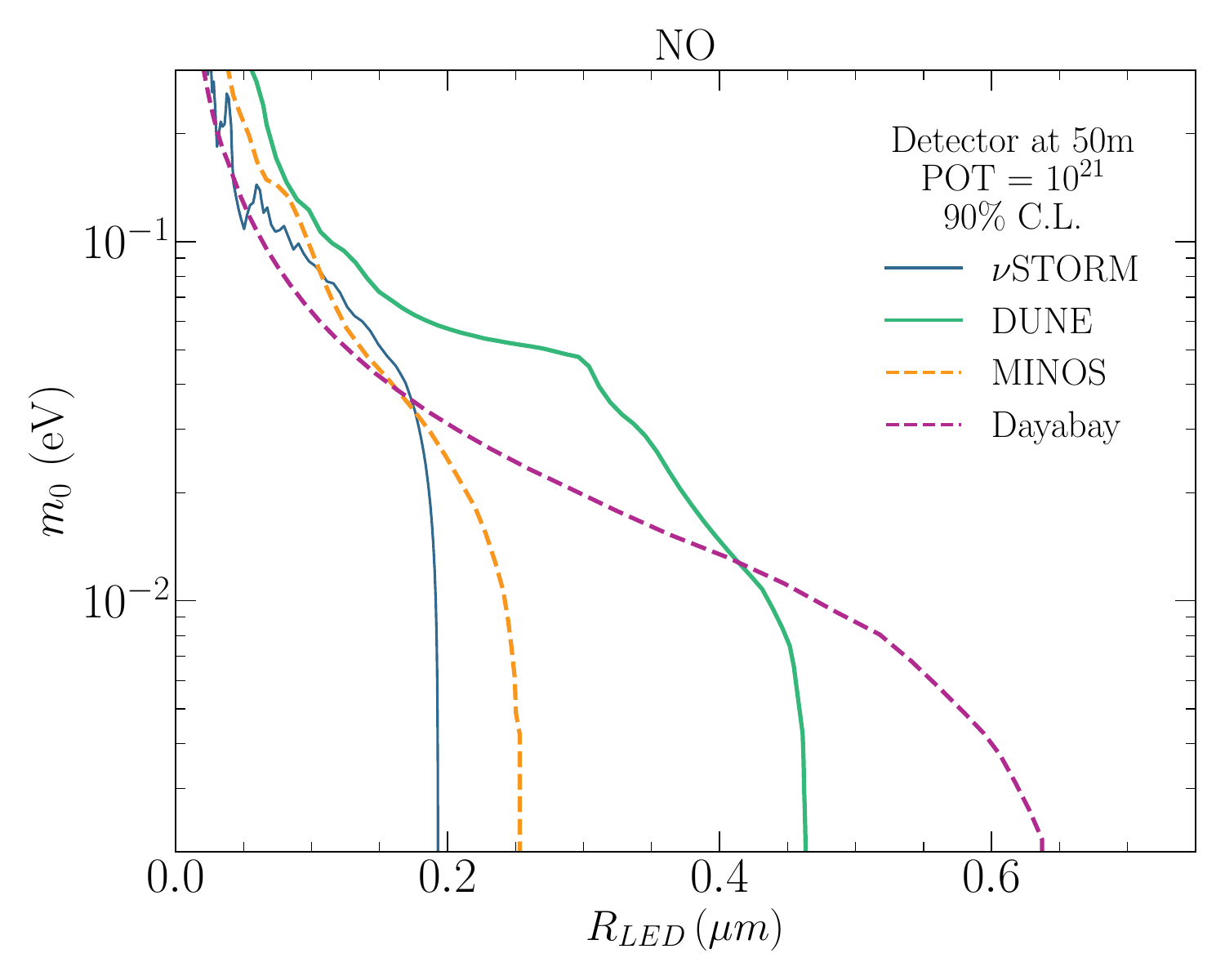}
        \caption{Chi-square contour Normal Ordering}
        \label{fig:chi_square_contour_LED_NO}
    \end{subfigure}
    \begin{subfigure}[b]{0.49\textwidth}
        \centering
        \includegraphics[width=1\linewidth]{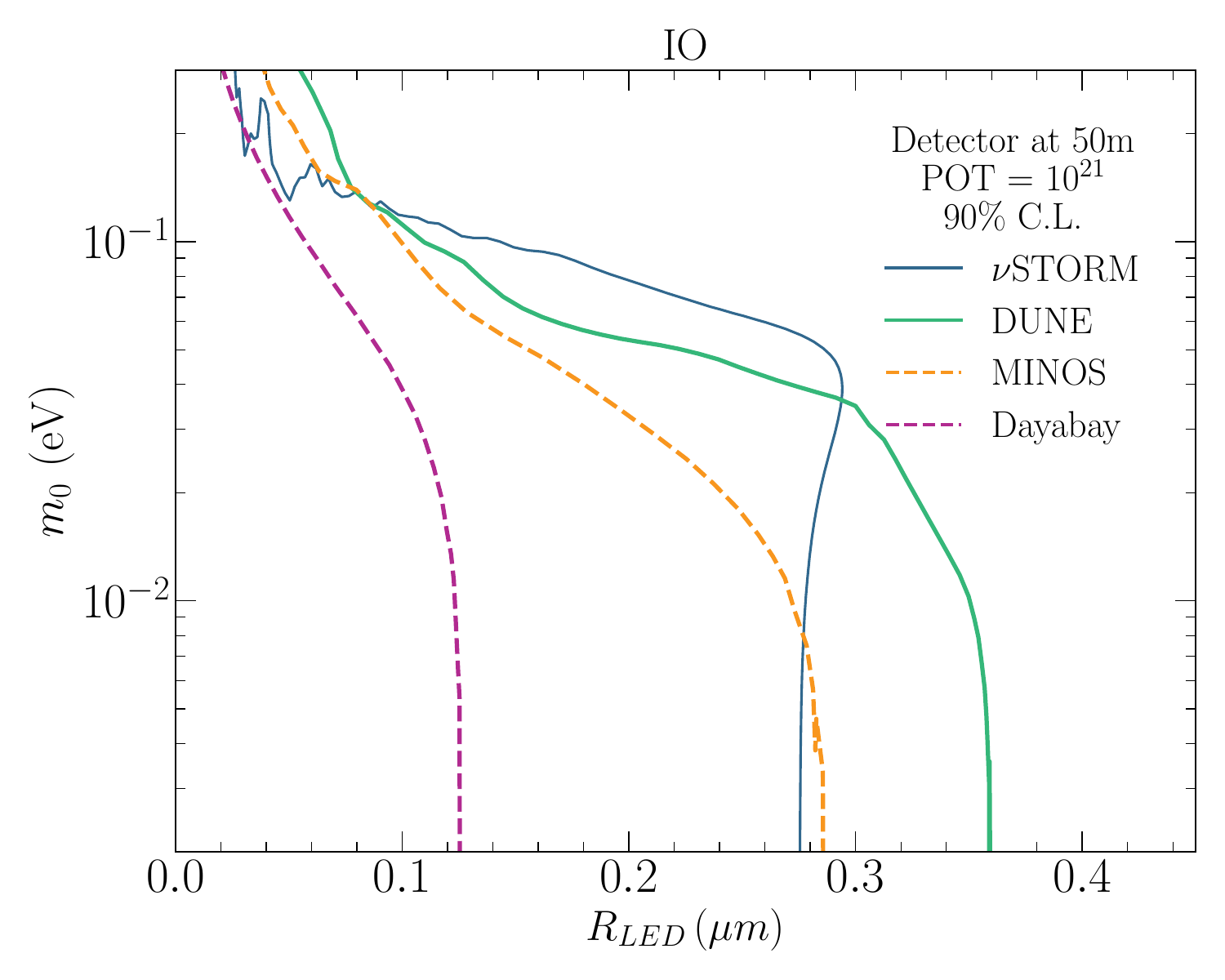}
        \caption{Chi-square contour Inverted Ordering}
        \label{fig:chi_square_contour_LED_IO}
    \end{subfigure}
        \caption{Sensitivity to disappearance under the assumption of 1\% systematic uncertainty. Comparison against contour from DUNE~\cite{Siyeon:2024pte}, MINOS, and Daya Bay~\cite{Forero:2022skg} is also shown. Bounds from existing measurements are shown with dashed lines, whereas bounds from projected constraints are shown in solid.}
    \label{fig:chi_square_led_oscillation} 
\end{figure}

\figref{fig:chi_square_led_oscillation} shows the projected $90\%$ C.L. exclusion contours in the $(R_{\mathrm{LED}}, m_0)$ plane for both normal and inverted mass ordering. In the NO scenario (\figref{fig:chi_square_contour_LED_NO}), nuSTORM achieves the most stringent bound among all experiments considered. This enhanced sensitivity arises from its intense and well-characterised $\nu_\mu$ beam from pion and muon decays. In the IO case (\figref{fig:chi_square_contour_LED_IO}), nuSTORM remains highly competitive, providing stronger constraints than DUNE and MINOS. The only experiment providing stronger constraints is Daya Bay, whose large $\nu_e$ data set benefits from the greater sensitivity of the $\nu_e \rightarrow \nu_e$ channel to LED effects in the IO regime. These results underscore the potential of nuSTORM to contribute competitively to the search for LED signatures, offering complementary coverage to existing long baseline and reactor experiments.


\subsection{Lepton Flavour Violation}\label{sec:LFV}
The discovery of neutrino oscillations confirmed that lepton flavour is not a conserved symmetry of nature. Within the Standard Model, however, flavour transitions between \emph{charged} leptons are suppressed by tiny neutrino masses to an unobservable level. Any measurable signal of charged lepton flavour violation (LFV) would therefore be a definitive sign of new physics~\cite{Kuno:1999jp,Bernstein:2013hba}. Many well-motivated frameworks beyond the Standard Model, such as supersymmetry or leptoquark models, predict LFV rates that could be within reach of modern experiments~\cite{Lindner:2016bgg,Calibbi:2017uvl}. Experimental searches have historically focused on rare muon processes like $\mu \rightarrow e\gamma$ and $\mu-e$ conversion in nuclei~\cite{MEG:2016leq,Bernstein:2019fyh}. Despite great effort, no evidence for LFV has been found, leading to stringent limits on these channels.

Light meson decays, the very source of the nuSTORM beam, offer a complementary avenue to search for LFV. The current best limit on the flavour-violating pion decay, $\mathrm{BR}(\pi^{+}\!\to\!\mu^{+}\nu_{e}) < 8\times10^{-3}$, was set by the BEBC experiment~\cite{Cooper-Sarkar:1981bam,Lyons:1981xs,ParticleDataGroup:2022pth} and several upcoming facilities aim to improve upon it~\cite{Alves:2024djc}. With its high-intensity and precisely understood beam, nuSTORM is in an excellent position to significantly tighten these bounds.
The specific LFV process of interest, $\pi^+ \rightarrow \mu^+ \nu_e$, would manifest as an anomalous excess of electron neutrinos in the detector. The expected number of signal events, $N_{\nu_e, \mathrm{sig}}$, is calculated by normalizing the observed number of standard pion decay events, $N_{\nu_\mu}^{\pi^+}$, with the LFV branching ratio:
\begin{equation}
N_{\nu_e, \mathrm{sig}} = N_{\nu_\mu}^{\pi^+} \cdot \mathrm{BR}(\pi^+ \rightarrow \mu^+ \nu_e)\,.
\end{equation}
The sensitivity is assessed using a simple $\chi^2$ statistic, assuming the signal-limited regime~\cite{Alves:2024djc}:
\begin{equation}
\chi^2 = \frac{N_{\nu_e, \mathrm{sig}}^2}{N_{\nu_e} + \sigma_{\nu_e}^2 N_{\nu_e}^2}\,,
\end{equation}
where $N_{\nu_e}$ is the expected intrinsic $\nu_e$ background and $\sigma_{\nu_e}$ is its associated systematic uncertainty.
\begin{table}[t!]
    \centering
    \begin{tabular}{l c}
        \toprule
        \multicolumn{2}{c}{\textbf{LFV (95\% C.L.)}} \\
        \midrule
        \textbf{Experiment (Uncertainty)} & $\mathrm{BR}\!\left(\pi^{+} \rightarrow \mu^{+} \nu_{e}\right)$ \\
        \midrule
        BEBC & $8.0 \times 10^{-3}$ \\
        SBND (10\%) & $1.5 \times 10^{-3}$ \\
        SBND-PRISM (10\%, 5\%) & $1.2 \times 10^{-3}$ \\
        SBND-PRISM (10\%, 2\%) & $8.9 \times 10^{-4}$ \\
        $\nu$STORM (1\%) & $7.1 \times 10^{-4}$ \\
        $\nu$STORM (Statistics only) & $4.7 \times 10^{-5}$ \\
        \bottomrule
    \end{tabular}
    \caption{Comparison of the bounds on LFV from BEBC~\cite{Cooper-Sarkar:1981bam,Lyons:1981xs,ParticleDataGroup:2022pth} and SBND-PRISM~\cite{Alves:2024djc}.}
    \label{Table: LFV BR comparision}
\end{table}
Table~\ref{Table: LFV BR comparision} summarises the projected 95\% C.L. upper limits on $\mathrm{BR}(\pi^+ \rightarrow \mu^+ \nu_e)$. nuSTORM is projected to set a bound of $7.1 \times 10^{-4}$, an improvement of more than an order of magnitude over the current limit from BEBC and roughly a factor of two beyond the best projection for SBND-PRISM. This powerful result is achieved despite nuSTORM having a higher intrinsic background ratio ($(\nu_e:\nu_{\mu}) \sim 0.036$) compared to BEBC ($\sim 0.016$)~\cite{Cooper-Sarkar:1981bam} and SBND ($\sim 0.008$)~\cite{Alves:2024djc}. The higher background is overcome by nuSTORM's immense statistical power, which will collect an order of magnitude more $\nu_\mu$ CC events ($\mathcal{O}(10^8)$) than other facilities (BEBC: $6060\pm440$, SBND: $\mathcal{O}(10^7)$).

In addition, the ability of $\nu$STORM to distinguish between $\nu_e$ and $\bar\nu_e$ allows a direct search for lepton number violating (LNV) in decays such as $\pi^+ \rightarrow \mu^+ \bar\nu_e$. This channel is essentially background-free, enabling sensitivity estimates based on the zero-background limit~\cite{Feldman:1997qc,ParticleDataGroup:2024cfk}: 
\begin{equation}
\mu_{95} = N_{\nu_{\mu}}^{\pi^+}\cdot \mathrm{BR}(\pi^+ \rightarrow \mu^+ \bar\nu_e)\,, 
\end{equation}
where $\mu_{95} \approx 3$ corresponds to the 95\% C.L. limit for a Poisson variable with no observed events, assuming a 1\% systematic uncertainty. Under these conditions, nuSTORM is projected to set an unprecedented bound of $\mathrm{BR}(\pi^+ \rightarrow \mu^+ \bar\nu_e) < 4.6\times 10^{-8}$, surpassing existing and projected limits from BEBC~\cite{Cooper-Sarkar:1981bam,Lyons:1981xs,ParticleDataGroup:2022pth} and SBND-PRISM~\cite{Alves:2024djc} of $\mathcal{O}(10^{-3})$~\cite{Alves:2024djc}. 

The improvement over BEBC is driven by the intense pion beam at $\nu$STORM, while the ability to distinguish between neutrino and antineutrino final states provides an advantage over SBND. While the intrinsic background is expected to be negligible, a careful study of potential misidentification of $\nu_e$ as $\bar\nu_e$ (and vice versa) will be required to robustly establish the achievable sensitivity.

 This enables nuSTORM to surpass previous and near-future bounds, demonstrating its unique capability in searches for rare processes.

\subsection{Heavy QCD Axions and ALPs}\label{sec:axion}

The Standard Model Lagrangian permits a CP-violating term in the QCD sector,
\begin{equation}
    \label{eq:qcd-theta}
    \mathcal{L}_\theta = - \frac{\theta g_s^2}{32\pi^2}\, G_{\mu\nu}^a \tilde{G}^{\mu\nu\,a}\,,
\end{equation}
parameterised by an angle $\bar{\theta} = \theta + \text{arg}\,\det[Y_u Y_d]$. Experimental bounds on the neutron electric dipole moment constrain this angle to be extraordinarily small, $|\bar{\theta}| \lesssim 10^{-10}$, a fine-tuning known as the Strong CP Problem~\cite{Wurm:2019yfj,Pendlebury:2015lrz}.

The Peccei–Quinn (PQ) mechanism~\cite{Peccei:1977hh} provides a dynamical solution by promoting $\bar{\theta}$ to a field, the axion $a(x)$. This is achieved by introducing a global U(1)$_{\text{PQ}}$ symmetry that is spontaneously broken at a high-energy scale, $f_a$. The axion is the pseudo-Nambu–Goldstone boson of this symmetry breaking. Its coupling to gluons is given by:
\begin{equation}
    \label{eq:axion-coupling}
    \mathcal{L}_a = \frac{a}{f_a} \frac{g_s^2}{32\pi^2} G_{\mu\nu}^a \tilde{G}^{\mu\nu\,a}\,,
\end{equation}
where $g_s$ is the QCD gauge coupling and $G_{\mu\nu}^a$ is the gluon field-strength tensor with colour index $a$.
This term introduces a potential which dynamically relaxes the effective CP-violating angle to zero. Non-perturbative QCD effects generate a mass for the axion that is inversely proportional to the breaking scale,
\begin{equation}
    m_a \approx 5.7~\mathrm{meV} \left(\frac{10^9~\mathrm{GeV}}{f_a}\right)\,.
\end{equation}
In the standard ``invisible axion'' scenario, astrophysical~\cite{Ayala:2014pea,Raffelt:1985nj,Blinnikov:1994eoa} and cosmological~\cite{Raffelt:1990yz,Raffelt:2006cw,Bar:2019ifz} constraints force $f_a$ to be very large ($ \gtrsim 10^8~\text{GeV}$), implying that the QCD axion must be extremely light ($m_a \lesssim 100~\text{meV}$) and feebly coupled, motivating searches with haloscopes~\cite{ADMX:2018ogs} and helioscopes~\cite{CAST:2013bqn,Irastorza:1567109}.

A significant theoretical challenge to this picture is the ``axion quality problem''. Global symmetries like the PQ symmetry are expected to be broken by quantum gravity effects, described by higher-dimensional operators~\cite{Co:2022bqq,Kelly:2020dda,Holman:1992us,Ghigna:1992iv,Barr:1992qq}, which would reintroduce the Strong CP Problem. This motivates models where the PQ symmetry is protected or where the axion's properties are modified. Several such frameworks have been proposed that allow for ``heavy'' QCD axions in the MeV-GeV mass range. In these models, the axion mass receives contributions from new physics (such as a dark confining sector or a mirror Standard Model~\cite{Co:2022bqq,Hook:2019qoh,Berezhiani:2000gh}), modifying the standard mass-to-decay-constant relation. Consequently, the decay constant $f_a$ can be in the TeV range while the axion mass is kinematically accessible at accelerators, providing much larger couplings to SM particles and opening new avenues for discovery.

The presence of kaons at nuSTORM offers a valuable opportunity to probe rare decay channels. A key feature of the facility is that undecayed kaons are absorbed at the end of the production straight, ensuring that searches involving kaons can be pursued without introducing backgrounds to other physics programs. Furthermore, the variable momentum control at nuSTORM allows one to select an appropriate central momentum to optimise the sensitivity of such studies. In this work, we assume a central momentum of $6.8\,\text{GeV}/c$ for the kaon beam, with an intensity of $\mathcal{O}(10^{17})$ kaons per $10^{21}$ protons on target at the beginning of the production straight. This choice of momentum balances kaon survival through the production straight with the kinematic reach required to maximise sensitivity to rare decay signatures.

The low-energy effective Lagrangian for a heavy QCD axion includes derivative couplings to quarks ($q$) and lepton ($\ell$) fields:
\begin{equation}
    \label{eq:axion-fermion-lagrangian}
    \mathcal{L}_{\text{int}} =
    \frac{C_q}{2 f_a}\, (\partial_\mu a)\, \bar{q}\gamma^\mu\gamma_5 q
    + \frac{C_\ell}{2 f_a}\, (\partial_\mu a)\, \bar{\ell}\gamma^\mu\gamma_5 \ell\,,
\end{equation}
where $C_q$ and $C_\ell$ are dimensionless model-dependent couplings to quarks  and charged leptons respectively. 
This class of models is crucial for nuSTORM, which can search for axions produced in rare meson decays, specifically $K^+ \rightarrow \pi^+ + a$, followed by the axion's visible decay, $a \rightarrow \mu^+ \mu^-$. The branching ratio for kaon decay to an axion is given by~\cite{Bauer:2021wjo,Bauer:2021mvw}
\begin{equation}
    \frac{\operatorname{Br}\left(K^{ \pm} \rightarrow a+\pi^{ \pm}\right)}{\operatorname{Br}\left(K_S^0 \rightarrow \pi^{+} \pi^{-}\right)}=
    \frac{\tau_{K^{ \pm}}}{\tau_{K_S}}
    \frac{2 f_\pi^2 c_3^2}{f_a^2}
    \left(
    \frac{m_K^2-m_a^2}{4 m_K^2-3 m_a^2-m_\pi^2}
    \right)^2
    \sqrt{
    \frac{
    \lambda\left(1, m_\pi^2 / m_K^2, m_a^2 / m_K^2\right)
    }{
    1-4 m_\pi^2 / m_K^2
    }}\,,
    \label{eq:KtoApi}
\end{equation}
where $c_3$ is the axion–gluon coupling and $\lambda$ is the Källén function. The partial width for the axion decay into dimuons is
\begin{equation}
    \Gamma\left(a \rightarrow \mu^+ \mu^-\right)
    =
    \frac{c_\mu^2 \, m_a \, m_\mu^2}{8 \pi f_a^2}
    \sqrt{1-\frac{4 m_\mu^2}{m_a^2}}\,,
    \label{eq:adecaywidth}
\end{equation}

where $c_\mu$ is the axion-muon coupling. The expected number of observable events is calculated as 
\begin{equation}
    N_{\text{obs}} = \Phi_K \cdot \operatorname{Br}(K \to a\pi) \cdot P_{\text{decay}} \cdot \epsilon_{\text{geom}}\,,
    \label{eq:naobs}
\end{equation}
where \( \Phi_K \) is the integrated kaon flux, \( \operatorname{Br}(K \to a\pi) \) is the production branching ratio, \( P_{\text{decay}} \) is the probability for the axion to decay within the detector and \( \epsilon_{\text{geom}} \) is the geometric acceptance. The decay probability depends on the axion's decay length in the lab frame, $L_a = (p_a/m_a) / \Gamma_{a \to \mu\mu}$ and the detector geometry:
\begin{equation}
    P_{\text{decay}}
    =
    e^{-L/L_a} \left(1 - e^{-\Delta L/L_a}\right).
    \label{eq:decayprob}
\end{equation}

\begin{figure}[t]
    \centering
    \includegraphics[width=0.49\linewidth]{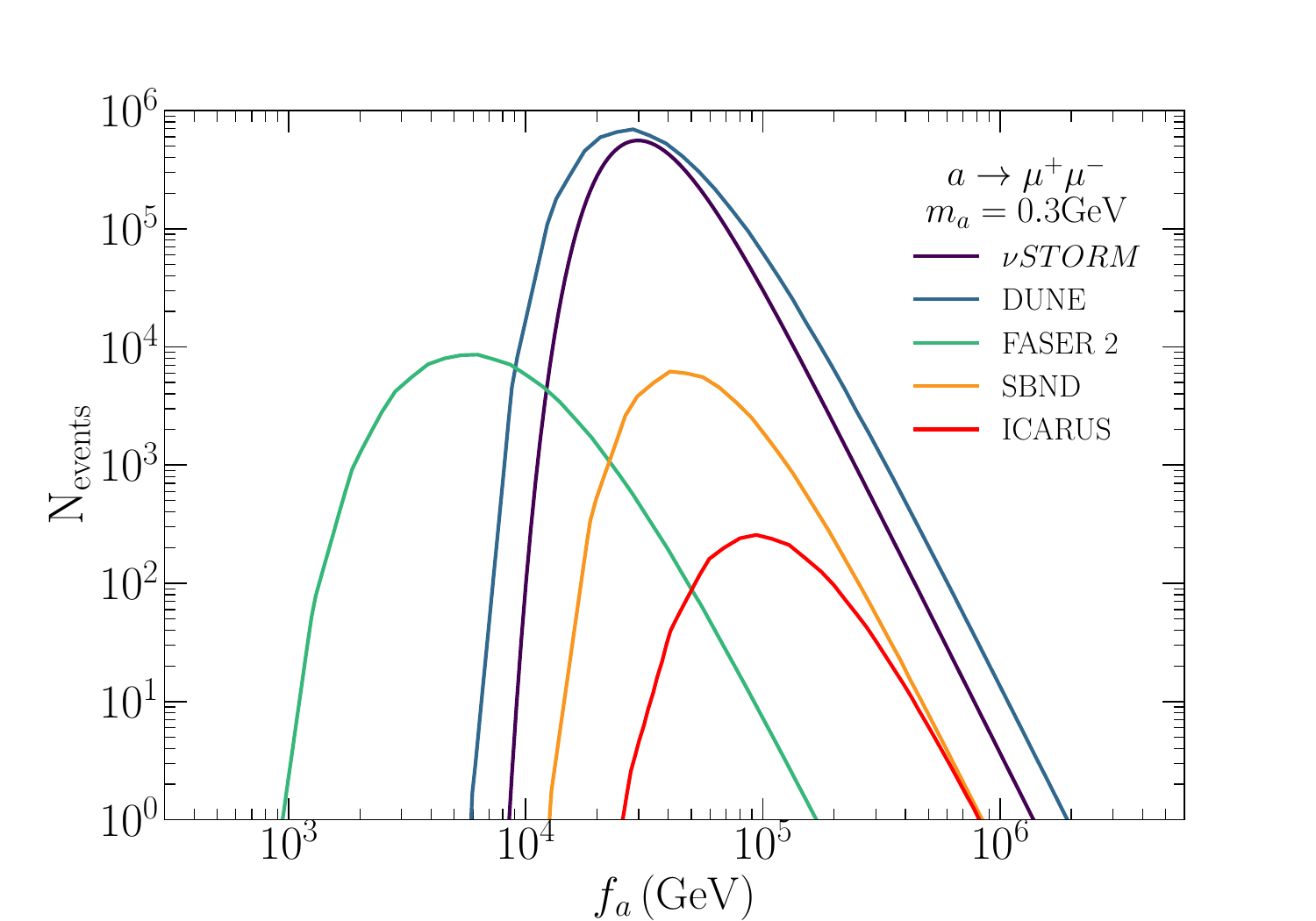}
    \includegraphics[width=0.49\linewidth]{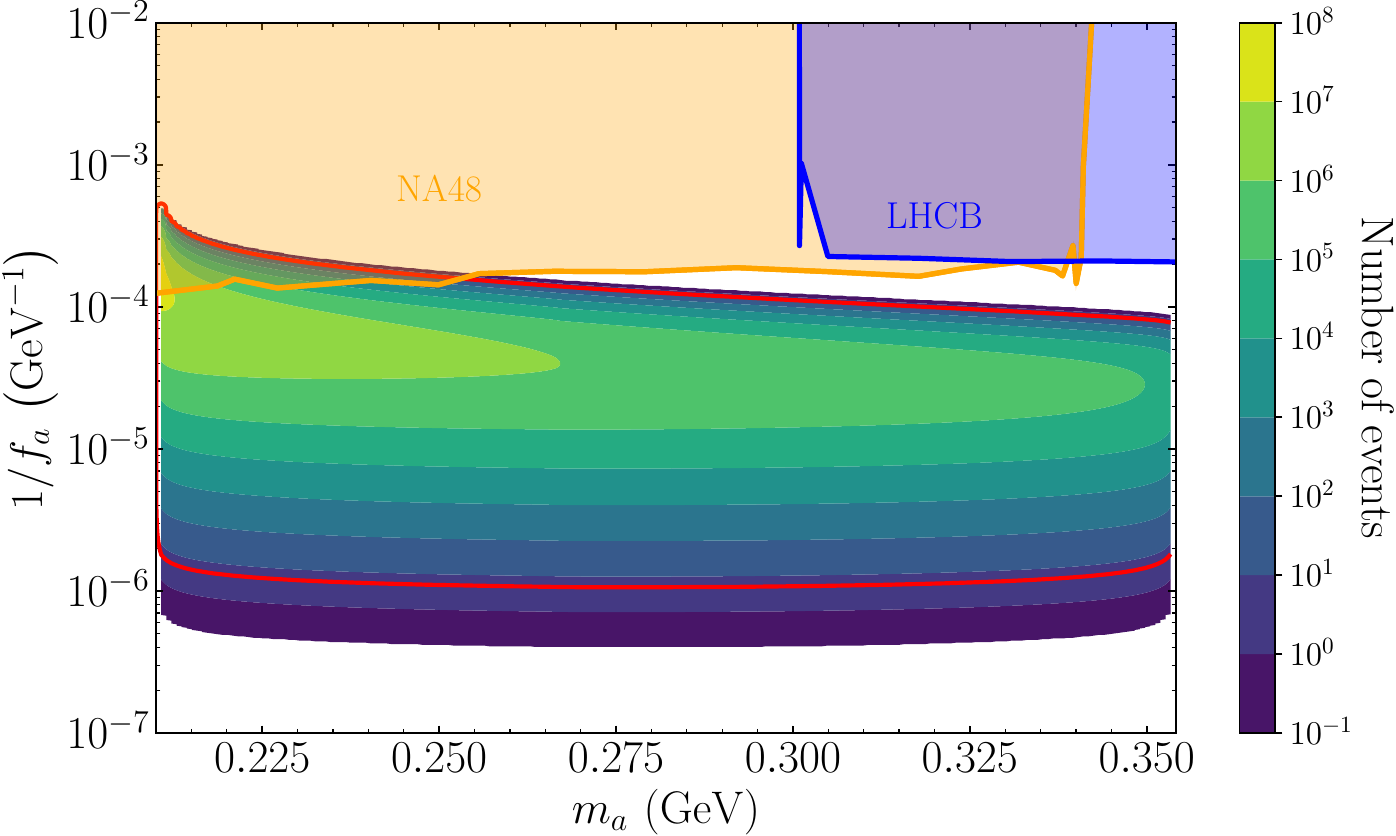}
    \caption{Expected number of events where axions decay into dimuons within the detector volume as a function of the decay constant $f_a$. An illustrative choice of $m_a = 0.3 \mathrm{GeV}$ and $c_l=1/36$ is made. }
    \label{fig:axion number of events}
\end{figure}

Given the highly boosted kaons at nuSTORM, the resulting axions are produced nearly collinear to the beam direction. We can therefore assume a geometric acceptance $\epsilon_{\text{geom}} \approx 100\%$. Furthermore, for the axion mass range considered, the branching ratio $\text{BR}(a \rightarrow \mu^+ \mu^-)$ is approximately unity, which simplifies the analysis~\cite{ArgoNeuT:2022mrm}. Although backgrounds for these rare decays are expected to be low, they are not entirely absent and may arise, for example, from neutrino trident production (\secref{sec:trident}). Despite differences in kinematical signatures, some overlap between the two sectors is possible. For the purposes of this study, we adopt an optimistic projection and assume no significant backgrounds. Nevertheless, given the sensitivity of short-baseline experiments, a dedicated background study is motivated to enable definitive probes of the hidden sector.
We compare our projected event rate to that of existing and proposed short-baseline neutrino experiments. 

The left panel of \figref{fig:axion number of events} shows the expected number of signal events for a benchmark axion mass of $m_a = 0.3~\mathrm{GeV}$ at the relevant exposures. Under these conditions, nuSTORM exhibits a significantly larger event rate than SBND and ICARUS. While FASER~2 may probe lower $f_a$ values, this region is already constrained by existing searches at LHCb~\cite{LHCb:2015nkv} and astrophysical limits such as those from NA48~\cite{NA482:2016sfh}. The right panel of \figref{fig:axion number of events} shows the projected reach in the $(m_a, f_a)$ parameter space, assuming a minimum of 5 events for discovery. nuSTORM is sensitive to a compelling region of parameter space, covering $f_a \sim \mathcal{O}(10^{-4}$-$10^{-6})~\mathrm{GeV}$ over the mass range $2 m_\mu \lesssim m_a \lesssim m_K - m_\pi$ and demonstrates competitive sensitivity relative to DUNE. The mass range is determined by the kinematics of axion production from $K^{\pm}\rightarrow \pi^{\pm} + a$. Constraints from ArgoNeuT~\cite{ArgoNeuT:2022mrm} further probe higher $f_a \simeq 10^{-4}~\mathrm{GeV}$ and nuSTORM, with its short baseline, could provide complementary coverage down to $f_a \simeq 10^{-6}~\mathrm{GeV}$. These techniques could also be extended to a variety of other dark-sector models, opening avenues for probing hidden particles. 

\section{Summary and Outlook}\label{sec:conclusion}
The primary strength of nuSTORM lies in its well-understood beam, derived from captured-pion and stored muon decays. Either source delivers a precisely known mixture of neutrino flavours. The composition is well defined and uniquely includes both $\nu_e$ and $\nu_{\mu}$. With a capable detector, this is a major advantage. Together with flux normalisation understood to better than 1\% through machine monitoring, systematic uncertainties on absolute rate predictions and cross-section measurements can be reduced to the percent level. By enabling high-statistics measurements across a broad range of channels, nuSTORM can address pressing systematic issues in neutrino physics while providing fertile ground for the discovery of phenomena beyond the Standard Model.

We have shown that nuSTORM offers strong sensitivity to sterile neutrinos in the $\Delta m^2_{41} \sim \mathcal{O}(1\,\text{eV}^2)$ regime, which is consistent to the analysis previously performed in \cite{nuSTORM:2014phr,Adey:2015iha}. Its ability to probe both $\nu_e$ and $\bar{\nu}_\mu$ disappearance channels with excellent control over systematics allows it to significantly improve upon existing global constraints.
Our analysis demonstrates that nuSTORM also has competitive sensitivity to Large Extra Dimensions (LED), setting projected exclusion bounds on the compactification radius of $R_{\text{LED}} \lesssim 0.1\,\mu\text{m}$ for the lightest mass state $m_0 \gtrsim 0.1\,\text{eV}$. This reach exceeds that of current and near-future experiments like DUNE~\cite{Siyeon:2024pte}, MINOS and Daya Bay~\cite{Forero:2022skg} in this mass regime.
For the weak mixing angle, a parameter of central importance to the Standard Model, nuSTORM enables statistically robust measurements using both $\bar{\nu}_\mu$ and $\nu_e$ scattering. These channels allow a measurement with percent-level precision under realistic systematic assumptions. 
In the search for lepton flavour violation (LFV), nuSTORM is projected to significantly improve upon existing constraints from BEBC~\cite{Cooper-Sarkar:1981bam,Lyons:1981xs,ParticleDataGroup:2022pth} and future bounds from SBND-PRISM~\cite{Alves:2024djc}. By probing $\pi^+ \to \mu^+ \nu_e$ decays, limits of $\text{BR} < 7.1 \times 10^{-4}$ at 95\% C.L. can be achieved, with the potential for even stronger constraints in a statistics-limited scenario. Similarly, the $\pi^+ \to \mu^+ \bar{\nu}_e$ channel provides a clean probe of lepton number violation (LNV), with an expected sensitivity of $\text{BR} < 4.6 \times 10^{-8}$ at 95\% C.L., exceeding the projected reach of SBND-PRISM~\cite{Alves:2024djc} as well as BEBC~\cite{Cooper-Sarkar:1981bam,Lyons:1981xs,ParticleDataGroup:2022pth}.

Furthermore, nuSTORM's high flux makes it an ideal platform for studying neutrino trident production. Our analysis indicates that nuSTORM could detect a substantial number of both coherent and diffractive trident events, with a reach competitive with DUNE and far exceeding that of current-generation experiments~\cite{Ballett:2018uuc}. The major background sources for these processes come from kaon decays at nuSTORM~\cite{Adey:2015iha}, although they provide negligible backgrounds as the pion flux is roughly two orders of magnitude larger than that of the kaon. However, with the additional access to kaons at nuSTORM, our exploratory analysis shows that nuSTORM has the potential to probe novel regions of the parameter space for axions and axion-like particles (ALPs) produced in kaon decays and subsequently decaying to dimuon final states, offering strong sensitivity in complement to collider~\cite{LHCb:2015nkv} and astrophysical~\cite{NA482:2016sfh} searches. This study underscores nuSTORM's versatility in testing for weakly coupled new physics.

The timely realisation of nuSTORM would substantially advance the global particle physics programme by providing essential cross-section measurements for long-baseline experiments and opening unique windows to new physics. Furthermore, its construction is timely, representing a critical and synergistic step on the path towards a Muon Collider.

\section*{Acknowledgments}
We would like to thank our fellow nuSTORM collaborators, in particular, Peter Hobson, Paul Kyberd, Paul Bogdan Jurj, Xianguo Lu, and Stefania Ricciardi, for their helpful comments and thorough proofreading. We would also like to thank Paul Kyberd and Paul Bogdan Jurj for their indispensable support with the neutrino fluxes.
J.T. would like to thank the Quantum Field Theory Centre at the University of Southern Denmark for their hospitality during the completion of this work.
YFPG has been supported by the Consolidaci\'on Investigadora grant CNS2023-144536 from the Spanish Ministerio de Ciencia e Innovaci\'on (MCIN) and by the Spanish Research Agency (Agencia Estatal de Investigaci\'on) through the grant IFT Centro de Excelencia Severo Ochoa No CEX2020-001007-S.

\bibliographystyle{JHEP}
\bibliography{biblio}

\end{document}